\documentclass[prc,aps,float,nofootinbib,superscriptaddress,twocolumn]{revtex4-1}
\usepackage{amssymb}
\usepackage{amsmath}
\usepackage{amsfonts}
\usepackage{epsfig}
\usepackage{bbm}
\usepackage{comment}
\usepackage{multirow}
\usepackage{longtable}
\usepackage{array}
\usepackage{subfigure}
\usepackage{graphics}
\usepackage{array}
\usepackage[usenames,dvipsnames]{color}
\allowdisplaybreaks

\begin{document}
\title{Ab initio calculation of the potential bubble nucleus $^{34}$Si}
\author{T. Duguet}
\email{thomas.duguet@cea.fr} 
\affiliation{IRFU, CEA, Universit\'e Paris-Saclay, 91191 Gif-sur-Yvette, France} 
\affiliation{KU Leuven, Instituut voor Kern- en Stralingsfysica, 3001 Leuven, Belgium}
\affiliation{National Superconducting Cyclotron Laboratory and Department of Physics and Astronomy, Michigan State University, East Lansing, MI 48824, USA}
\author{V. Som\`a}
\email{vittorio.soma@cea.fr} 
\affiliation{IRFU, CEA, Universit\'e Paris-Saclay, 91191 Gif-sur-Yvette, France} 

\author{S. Lecluse}
\email{simon.lecluse@student.kuleuven.be} 
\affiliation{KU Leuven, Instituut voor Kern- en Stralingsfysica, 3001 Leuven, Belgium}

\author{C. Barbieri}
\email{c.barbieri@surrey.ac.uk}
\affiliation{Department of Physics, University of Surrey, Guildford GU2 7XH, UK}

\author{P. Navr\'atil}
\email{navratil@triumf.ca}
\affiliation{TRIUMF, 4004 Westbrook Mall, Vancouver, BC, V6T 2A3, Canada}

\date{\today}

\begin{abstract} 
\begin{description}
\item[Background]  The possibility that an unconventional depletion (referred to as ``bubble'') occurs in the center of the charge density distribution of certain nuclei due to a purely quantum mechanical effect has attracted theoretical and experimental attention in recent years. Based on a mean-field rationale, a correlation between the occurrence of such a semi-bubble and an anomalously weak splitting between low angular-momentum spin-orbit partners has been further conjectured. Energy density functional and valence-space shell model calculations have been performed to identify and characterize the best candidates, among which $^{34}$Si appears as a particularly interesting case. While the experimental determination of the charge density distribution of the unstable $^{34}$Si is currently out of reach, (d,p) experiments on this nucleus have been performed recently to test the correlation between the presence of a bubble and an anomalously weak $1/2^{-}\!-\!3/2^{-}$ splitting in the spectrum of $^{35}$Si as compared to $^{37}$S.
\item[Purpose] We study the potential bubble structure of $^{34}$Si on the basis of the state-of-the-art ab initio self-consistent Green's function many-body method.
\item[Methods] We perform the first ab initio calculations of $^{34}$Si and $^{36}$S. In addition to binding energies, the first observable of interest are the charge density distribution and the charge root mean square radius for which experimental data exist in $^{36}$S. The second observable of interest is the low-lying spectroscopy of $^{35}$Si and $^{37}$S obtained from (d,p) experiments along with the spectroscopy of $^{33}$Al and $^{35}$P obtained from knock-out experiments. The interpretation in terms of the evolution of the underlying shell structure is also provided. The study is repeated using several chiral effective field theory Hamiltonians as a way to test the robustness of the results with respect to input inter-nucleon interactions. The convergence of the results with respect to the truncation of the many-body expansion, i.e. with respect to the many-body correlations included in the calculation, is studied in detail. We eventually compare our predictions to state-of-the-art multi-reference energy density functional and shell model calculations.
\item[Results] The prediction regarding the (non)existence of the bubble structure in $^{34}$Si varies significantly with the nuclear Hamiltonian used. However, demanding that the experimental charge density distribution and the root mean square radius of $^{36}$S are well reproduced,  along with $^{34}$Si and $^{36}$S binding energies, only leaves the NNLO$_{\text{sat}}$ Hamiltonian as a serious candidate to perform this prediction. In this context, a bubble structure, whose fingerprint should be visible in an electron scattering experiment of $^{34}$Si, is predicted. Furthermore, a clear correlation is established between the occurrence of the bubble structure and the weakening of the $1/2^{-}\!-\!3/2^{-}$ splitting in the spectrum of $^{35}$Si as compared to $^{37}$S.
\item[Conclusions] The occurrence of a bubble structure in the charge distribution of $^{34}$Si is convincingly established on the basis of state-of-the-art ab initio calculations. This prediction will have to be revisited in the future when improved chiral nuclear Hamiltonians are constructed. On the experimental side, present results act as a strong motivation to measure the charge density distribution of $^{34}$Si in future electron scattering experiments on unstable nuclei. In the meantime, it is of interest to perform one-neutron removal on $^{34}$Si and $^{36}$S in order to further test our theoretical spectral strength distributions over a wide energy range.

\end{description}
\end{abstract}
\maketitle

\section{Introduction}
\label{introduction}

The possibility that an unconventional depletion, referred to as ``bubble'', occurs in the center of the point-nucleon and/or charge density distributions of a nucleus has attracted theoretical as well as experimental attention in recent years. Earlier, so-called (semi-)bubble structures had been invoked mainly in connection with hypothetical (super-) hyper-heavy nuclei characterized by a very large charge ($120\leq Z\leq 240$) $240\leq Z\leq 280$~\cite{decharge03a}. Indeed, single-reference (SR) energy density functional (EDF) calculations predicted that the ground-state configuration of these nuclei may display a depletion in the center of their density distribution~\cite{decharge03a,bender_she99} as a result of a collective quantum mechanical effect sustained by the compromise between the large repulsive Coulomb interaction and the strong force that binds them~\cite{bender13a}.  

The case of present interest relates to nuclei characterized by more conventional masses and possibly displaying a semi bubble\footnote{As for the terminology, we use indistinguishably "bubble", "semi bubble" or "central depletion" in the following although strictly speaking "bubble" should be kept for speculative hyperheavy nuclei possibly displaying a null density in their center, which is not the case for the nucleus of present interest.} in their center. The rationale here solely relates to the quantum mechanical effect that finds its source in the sequence of occupied and unoccupied single-particle states near the Fermi energy in an independent-particle or a mean-field picture. While $s$ ($\ell =0$) orbitals display a radial distribution that is peaked at the center of the nucleus, orbitals with non-zero angular momenta ($\ell \neq0$) are suppressed in the nuclear interior such that they do not contribute to the central density. As a result, any vacancy of $s$ orbitals embedded among larger $\ell $ orbitals near the Fermi level is expected to produce a depletion of the central density. These hypothetical nuclei are of interest as they must be modelled via mean-field potentials that differ from those associated with Fermi-type density distributions that fit the vast majority of nuclei. In turn, a non-zero density derivative in the nuclear interior has been conjectured to cause a sharp increase of "non-natural" sign of the effective one-body spin-orbit potential, eventually inducing a reduction of the splitting between spin-orbit partners characterized by low angular momenta~\cite{ToddRutel04,burgunderthesis}. 

Going beyond this mean-field scenario, a small energy difference between the unoccupied $s$ shell and the last occupied/next unoccupied shells can favor collective correlations and thus lower or even wash out the depletion at the center of the potential bubble nucleus. Therefore the search for the best bubble candidates must be oriented towards nuclei that can be reasonably modelled by an $s$ orbital well separated from nearby single-particle states such that correlations are weak. In turn, this feature underlines the necessity to employ theoretical methods that explicitly incorporate long-range correlations modifying the density on a length scale of about $1$ fm, which is the typical expected spatial extent of the depletion at the center of bubble nuclei as discussed below.

In recent years, SR~\cite{richter03a,ToddRutel04,Khan07,Grasso09} and multi-reference (MR)~\cite{Yao12,Yao13,Wu14} EDF calculations along with shell model calculations~\cite{Grasso09} have been performed for $^{22}$O, $^{34}$Si, $^{46}$Ar, $^{204}$Hg and $^{206}$Hg. Indeed, these nuclei appeared as favorable candidates based on the naive filling of single-particle shells their number of protons and/or neutrons correspond to. Among those, $^{34}$Si ($Z=14$, $N=20$) stands out as the most viable case as its depletion factor defined as
\begin{equation}
F \equiv \frac{\rho_{{\rm max}}-\rho_{{\rm c}}}{\rho_{{\rm max}}}\, ,
\label{depletion_factor_bubble}
\end{equation}
is predicted to be the highest among all candidates in SR-EDF calculations. In Eq.~\ref{depletion_factor_bubble}, $\rho_{{\rm c}}$ and $\rho_{{\rm max}}$ denote central and maximum (point-nucleon or charge) density values, respectively. For $Z=14$, the naive filling of proton shells leaves the $1s_{1/2}$ single-particle state as the first unoccupied level above the Fermi energy. Furthermore, the $N=20$ magic character of $^{34}$Si translates into a first $2^+$ excitation energy ($E_{2^+_1}=3.3$\,MeV) and a $B(E2; 0^+_1\rightarrow 2^+_1)$  transition probability~\cite{ibbotson98a} that are similar to the doubly-magic $^{40}$Ca nucleus\footnote{See Ref.~\cite{Sorlin08a} and references therein for the systematic of $E_{2^+_1}$ and $B(E2)$ in the $N=20$ isotonic chain.}. The low electric monopole transition strength $\rho(E0; 0^+_1\rightarrow 0^+_2)$~\cite{rotaru12a} completes the picture of a doubly-magic system~\cite{baumann89a}. These features leaves the hope that the naive rationale based on an independent-particle model only needs to be slightly perturbed by the inclusion of long-range correlations\footnote{The perfect counter example is $^{28}$Si ($Z=14$, $N=14$) for which the even more promising picture provided by the naive filling of spherical shells is eventually fully invalidated, resulting in a charge density distribution that does not display any depletion in its center~\cite{miessen82a}.}.

A bubble structure mainly driven by protons, as in $^{34}$Si, can be probed directly by measuring the charge density distribution via electron scattering. However, it is presently not possible to perform electron scattering on unstable nuclei as light as $^{34}$Si with sufficient luminosity. Such an experiment may become feasible in the next decade at ELISe$@$FAIR~\cite{Simon07} or after an upgrade of the SCRIT facility at RIKEN~\cite{Suda09}. 

Because the presence of the central depletion is believed to correlate with specific quantum mechanical properties and to feedback on other observables, one may think of alternative ways to probe its presence indirectly, e.g. via direct reactions. In the present case of interest, we specifically wish to test the correlation between the presence of the bubble and the evolution of the $E^{+}_{1/2^-}-E^{+}_{3/2^-}$ spin-orbit splitting when going from $^{37}$S to $^{35}$Si. The establishment of this correlation is performed in the eye of the capacity of ab initio calculations to reproduce the low-lying spectroscopy of nuclei obtained via the addition of a neutron~\cite{Thorn84,Eckle89,Burgunder14} or the removal of a proton~\cite{khan85a,mutschler16a,mutschler16b} on $^{36}$S and $^{34}$Si. 

While potential bubble nuclei such as $^{34}$Si have  already been investigated quite thoroughly within the frame of EDF and shell-model many-body methods, the goal of the present work is to provide the first study based on ab initio many-body calculations. As mentioned above, our aim is to perform a coherent analysis of both density distributions and one-neutron addition and removal spectral strength distributions. Ideally, one would like to further correlate these observables with spectroscopic information in $^{34}$Si itself as was done in Refs.~\cite{Yao12,Yao13}. However, the many-body scheme employed  does not allow to do it yet. This will hopefully become possible in a not too distant future. Also, one of the objectives of the present study is to characterize the sensitivity of the results to the input Hamiltonian and to outline the role of three-nucleon forces.

The paper is organized as follows. Section~\ref{compute} focuses on the computational scheme employed, paying particular attention to the convergence of the observables of interest with respect to the basis used to represent the Schr\"odinger equation and to the many-body truncation implemented to solve it. Section~\ref{chap_dens} analyses in detail the characteristic of point-proton and charge density distributions of $^{36}$S and $^{34}$Si. The impact of many-body correlations and the sensitivity of the results to the utilized Hamiltonian are discussed. Results from our ab initio calculations are further compared to those obtained from state-of-the-art MR-EDF and SM calculations. Section~\ref{spectro} concentrates first on the reproduction of the spectroscopy of neighboring $^{37}$S, $^{35}$P, $^{35}$Si and $^{33}$Al. In particular we correlate the evolution of the $E^{+}_{1/2^-}-E^{+}_{3/2^-}$ spin-orbit splitting when going from $^{37}$S to $^{35}$Si and the presence/absence of a bubble structure in the ground state of $^{34}$Si/$^{36}$S. The interpretation in terms of the evolution of the underlying one-nucleon shell structure is also provided.

\section{Computational scheme}
\label{compute}

\subsection{Many-body methods and Hamiltonians}
\label{calcsetup}

The present analysis is based on self-consistent Green function (SCGF) calculations~\cite{Dickhoff:2004xx,Barbieri:2009nx,soma11a} of $^{34}$Si and $^{36}$S. This is done employing the following scheme
\begin{enumerate}
\item By default, a combination of two-nucleon (2N) and three-nucleon (3N) interactions obtained within the frame of chiral effective field theory ($\chi$EFT) at next-to-next-to leading order (N$^{2}$LO) and denoted as NNLO$_{\text{sat}}$~\cite{Ekstrom15} is used. For comparison, we occasionally employ a different set of N$^{2}$LO interactions~\cite{Ekstrom13}, denoted as NNLO$_{\text{opt}}$, 
as well as the combination of N$^3$LO 2N and N$^2$LO 3N chiral forces (NN+3N400) of Refs.~\cite{entem03, Navratil07, Roth12} evolved down to a low-momentum scale $\lambda=1.88 - 2.24\,\text{fm}^{-1}$ by means of a similarity renormalization group (SRG) transformation~\cite{Bogner:2009bt}. 
The comparison of the results based on the three sets of $\chi$EFT interactions is of interest given that NNLO$_{\text{sat}}$ was specifically designed~\cite{Ekstrom15} to address the impossibility of any existing set to  convincingly describe binding energies and nuclear radii (saturation) of mid-mass nuclei~\cite{Soma:2013xha, Hergert:2014iaa,Lapoux:2016exf} (infinite nuclear matter~\cite{Carbone14, Hag14nm}) at the same time. 
\item Calculations expand one-, two- and three-body operators on a spherical harmonic oscillator basis containing up to 14 harmonic oscillator (HO) shells [$N_{\text{{\rm max}}} \equiv$~max~$(2n+l)$ = 13]. All matrix elements of one- and two-body operators are used whereas those of 3N interactions are limited to configurations characterized by $N_1$+$N_2$+$N_3\leq N^{\text{3N}}_{\text{{\rm max}}}$=16.
\item As nuclei under study display a closed (sub-)shell character in a naive independent-particle approximation, both Dyson~\cite{Dickhoff:2004xx,Barbieri:2009ej,Cipollone:2013zma} and Gorkov~\cite{soma11a, Soma14a, Soma:2013xha} SCGF calculations can be safely performed. The latter framework is employed to test whether the explicit inclusion of pairing correlations via the breaking of U(1) global gauge symmetry associated with particle-number conservation modifies, e.g. improves, the theoretical predictions or not. The many-body truncation scheme is based on the so-called n$^{{\rm th}}$-order adiabatic diagrammatic construction (ADC(n))~\cite{Schirmer83}. While Gorkov SCGF calculations can currently be performed at ADC(1), i.e. Hartree-Fock(-Bogoliubov) (HF(B)), and second-order (i.e. ADC(2)) levels, Dyson SCGF further accesses ADC(3) calculations. While ADC(2) calculations already resum the bulk of dynamical correlations, ADC(3) provides well-converged bulk properties of the A-body system of interest along with a refined description of spectroscopic properties of A$\pm$1 nuclei.
\end{enumerate}

\subsection{Convergence of ground-state energies}

\begin{figure}[b]
\begin{center}
\includegraphics[width=8.5cm]{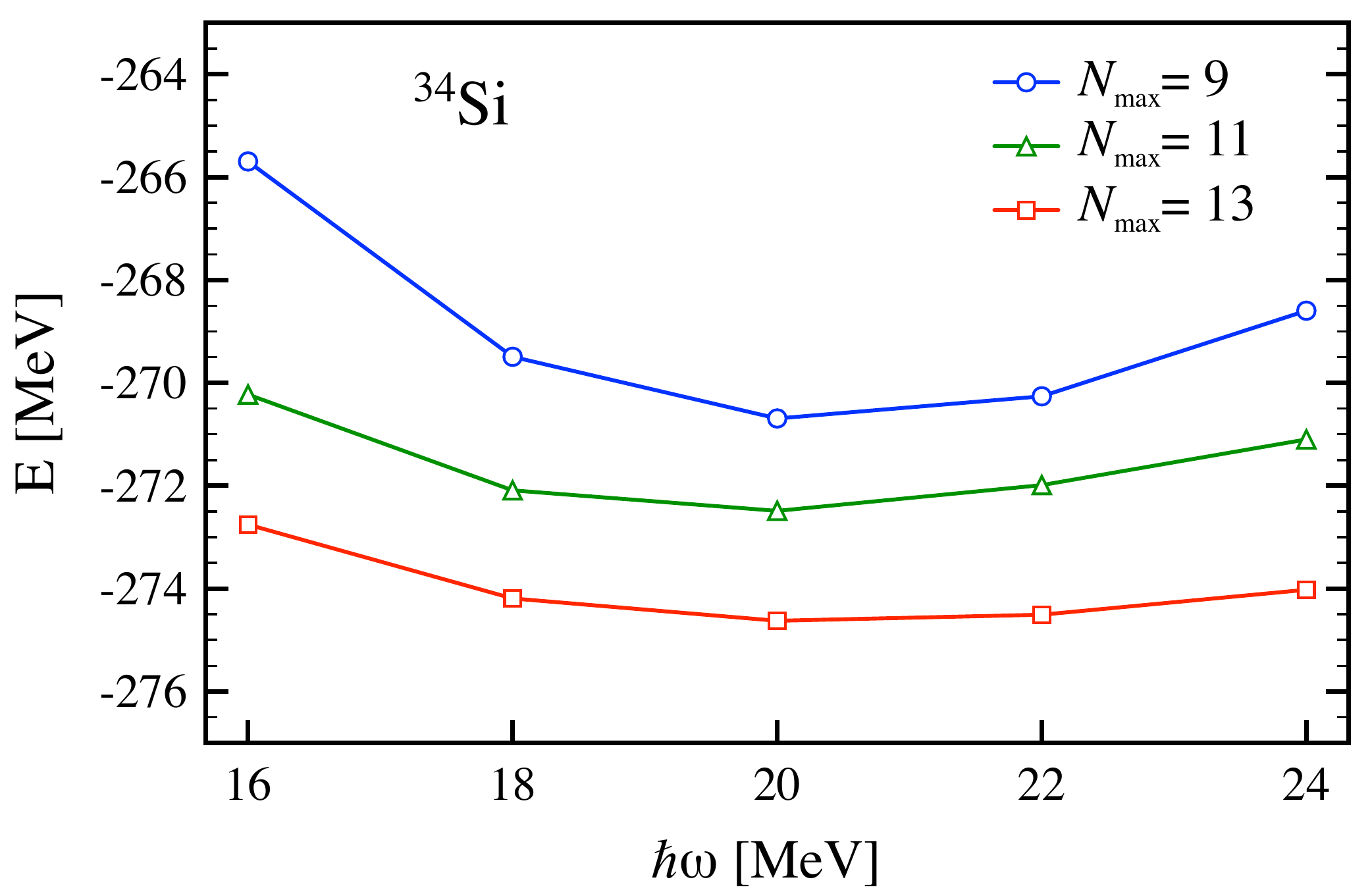}
\end{center}
\caption{(Color online) ADC(2) ground-state energy of $^{34}$Si as a function of the harmonic oscillator spacing $\hbar\omega$ and for increasing size $N_{\text{{\rm max}}}$ of the single-particle model space.}
\label{convergenceE}
\end{figure}

The ground-state energy of $^{34}$Si computed at the ADC(2) level is displayed in Fig.~\ref{convergenceE} as a function of the oscillator frequency $\hbar\omega$ and for increasing values of $N_{\text{{\rm max}}}$. As the number of major shells increases, the energies become more independent of $\hbar\omega$ while their minima shift progressively towards lower values to eventually reach $\hbar\omega=20$\,MeV for $N_{\text{{\rm max}}}=13$. The behavior is similar for $^{36}$S. 

In principle, the choice of a specific harmonic oscillator frequency is arbitrary in the limit of very large $N_{\text{{\rm max}}}$ as many-body quantities must become independent of it. At workable values of $N_{\text{{\rm max}}}$, observables are only approximately independent of $\hbar\omega$ and the optimal value of the latter may differ from one observable to the other and from one eigenstate to the other. For example, lower values of $\hbar\omega$ than the one minimizing the ground-state energy might better approach the infinite-basis value for long-range, e.g. mean-square radii, operators~\cite{Shin16}.  As discussed next, however, in the present case there is little impact of the specific value of $\hbar\omega$ on density distributions, which constitute the focus of the present paper. Consequently, and given the lack of a well defined extrapolation procedure for density distributions, the value $\hbar\omega=20$\,MeV corresponding to the minimum of the energy for $N_{\text{{\rm max}}}=13$ is considered in the following sections.
\begin{table}
\centering
\begin{tabular}{|c||c|c|c||c|}
\hline
$E$ &  \hspace{.1cm} ADC(1) \hspace{.1cm} & \hspace{.1cm} ADC(2) \hspace{.1cm} &  \hspace{.1cm} ADC(3) \hspace{.1cm}  & Experiment \\
\hline
\hline
\hspace{.1cm} $^{34}$Si \hspace{.1cm} & -84.481 & -274.626 & -282.938 & -283.427 \\
$^{36}$S  & -90.007 & -296.060 & -305.767 & -308.714 \\
\hline
\end{tabular}
\caption{Ground-state energies (in MeV) computed within ADC(1), ADC(2) and ADC(3) approximations. Experimental data are from Ref.~\cite{AME12}.}
\label{BE}
\end{table}

Ground-state energies computed at various orders in the many-body truncation scheme are compared to experimental data in Tab.~\ref{BE}. At the ADC(2) level, theoretical results are within $4\%$ of experimental data, which is consistent with missing ADC(3) correlations and the intrinsic uncertainty of the input Hamiltonian~\cite{Lapoux:2016exf, Ekstrom15}. Going to  ADC(3) indeed brings about 8-10 MeV additional binding, which represents about $5\%$ of the correlation energy generated at the ADC(2) level. Extrapolating the pattern of reduction in the correlation energy added at each ADC(n) order, the ADC(3) results can be safely believed to be about 1-2 MeV (i.e. less than $1\%$) away from the fully converged values. With the presently used NNLO$_{\text{sat}}$ Hamiltonian, this happens to be of the order of the difference to experimental data.

\subsection{Convergence of ground-state radii}

\begin{figure}[t]
\begin{center}
\includegraphics[width=8.5cm]{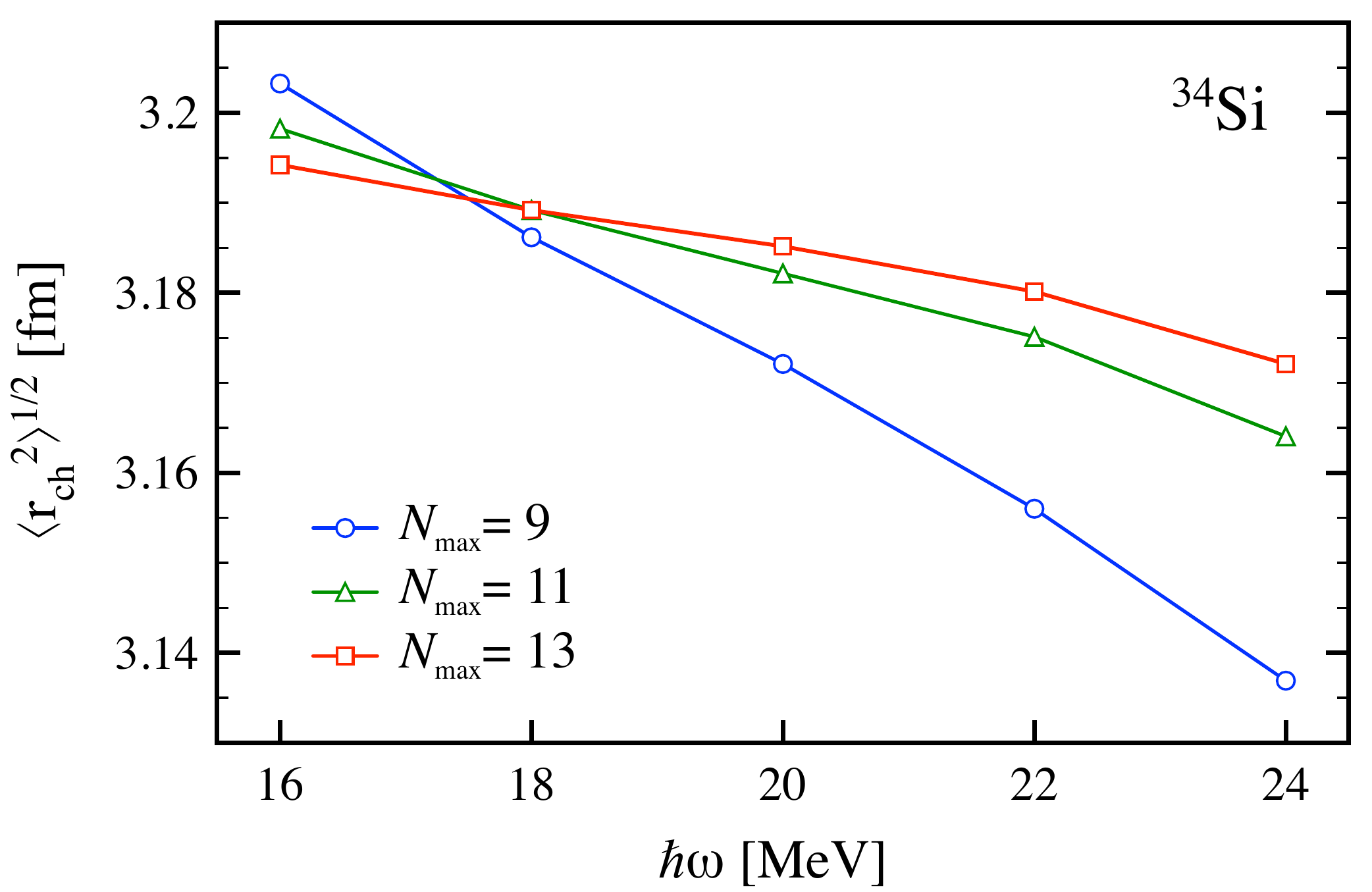}
\end{center}
\caption{(Color online) ADC(2) ground-state rms charge radius of $^{34}$Si as a function of the harmonic oscillator spacing $\hbar\omega$ and for increasing size $N_{\text{{\rm max}}}$ of the single-particle model space.}
\label{convergenceR}
\end{figure}

Before addressing point-nucleon and charge density distributions, let us focus on the integrated information constituted by point-nucleon and charge root-mean-square (rms) radii. In Fig.~\ref{convergenceR}, ADC(2) calculations of the charge rms radius\footnote{In the present work charge radii are computed from point-proton radii by accounting for the finite charge radii of both protons and neutrons in addition to the Darwin-Foldy correction, see Ref.~\cite{Cipollone15} for details.}
$\langle r^{2}_{{\rm ch}}\rangle^{1/2}$ of $^{34}$Si are displayed for different values of $\hbar\omega$ and $N_{\text{{\rm max}}}$. 
As $N_{\text{{\rm max}}}$ increases, the dependence on $\hbar\omega$ becomes weaker, totalling to about $2\%$ for $N_{\text{{\rm max}}}=13$ for $\hbar\omega \in [16,24]$\,MeV. 
Table~\ref{radiicorrelations} reports charge rms radii of $^{34}$Si and $^{36}$S computed within different many-body truncation schemes. 
The convergence pattern is similar for the two nuclei, with tiny differences between ADC(2) and ADC(3) results. This indicates that rms radii are essentially converged already at the ADC(2) level.

It is currently a challenge for ab initio calculations to describe both the binding energy and the size of medium-mass nuclei at the same time~\cite{Lapoux:2016exf}. This situation lead recently to the construction of the (unconventional) NNLO$_{\text{sat}}$ $\chi$EFT Hamiltonian~\cite{Ekstrom15} that is presently used and that indeed improves the situation significantly~\cite{Ruiz:2016gne,Lapoux:2016exf}. The computed value $\langle r^{2}_{{\rm ch}}\rangle^{1/2}=3.285$ fm in $^{36}$S is very close to the experimental measurement.
Comparatively, the rms charge radius computed from the NN+3N400 Hamiltonian processed through a SRG transformation is significantly too small, e.g. it is predicted to be $\langle r^{2}_{{\rm ch}}\rangle^{1/2}=2.867$ fm at the ADC(2) level for $\lambda=1.88$\,fm$^{-1}$.

Experimental charge radii are unavailable for the unstable $^{34}$Si nucleus. 
While charge radii for stable isotopes can be measured by means of electron scattering, collinear laser spectroscopy experiments~\cite{Campbell16} currently constitute the most appropriate way to access charge radii of unstable nuclei with lifetimes as low as a few milliseconds. 
However, Si elements are highly reactive and require a high evaporation temperature, thus are extremely difficult to produce and extract via ISOL techniques. 
In-flight facilities, e.g. NSCL at Michigan State University, are able to provide high-intensity beams of Si isotopes.
Future developments of high-resolution laser spectroscopy experiments should enable a measure of the rms charge radius of $^{34}$Si~\cite{ronald16a}. 
\begin{table}[t]
\centering
\begin{tabular}{|c||c|c|c||c|}
\hline
$\langle r^{2}_{{\rm ch}}\rangle^{1/2}$& ADC(1) & ADC(2) & ADC(3)  & Experiment \\
\hline
\hline
$^{34}$Si & 3.270 & 3.189 & 3.187 & - \\
$^{36}$S & 3.395 & 3.291 & 3.285 & 3.2985 $\pm$ 0.0024 \\
\hline
\end{tabular}
\caption{Charge rms radii (in fm) computed within ADC(1), ADC(2) and ADC(3) approximations. The experimental value is from Ref.~\cite{Angeli13}.}
\label{radiicorrelations}
\end{table}
\begin{table}
\centering
\begin{tabular}{|c||c|c|c|c|}
\hline
& $\langle r^{2}_{{\rm p}}\rangle^{1/2}$  & $\langle r^{2}_{{\rm n}}\rangle^{1/2}$  & $\langle r^{2}_{{\rm m}}\rangle^{1/2}$  & $\langle r^{2}_{{\rm ch}}\rangle^{1/2}$\\
\hline
\hline
\hspace{.1cm} $^{34}$Si \hspace{.1cm} & 3.085 & 3.258 & 3.188 & 3.187 \\
$^{36}$S  & 3.184 & 3.285 & 3.240 & 3.285 \\
\hline
\end{tabular}
\caption{Theoretical point-proton, point-neutron, matter and charge rms radii (in fm) calculated at the ADC(3) level.}
\label{tab:allRadii}
\end{table}

For completeness, point-proton, point-neutron, matter and charge radii computed at the ADC(3) level are reported in Tab.~\ref{tab:allRadii}. 
Recent works have demonstrated that matter radii can be reliably extracted from elastic proton scattering data~\cite{Lapoux:2015jva, Lapoux:2016exf}.
Given that such data are available for several sulfur isotopes including $^{36}$S~\cite{Marechal99}, it would be interesting to compare a similar evaluation of 
$\langle r^{2}_{{\rm m}}\rangle^{1/2}$ to our present results.

\subsection{Convergence of point-nucleon densities}

\begin{figure}[b]
\begin{center}
\includegraphics[width=8.7cm]{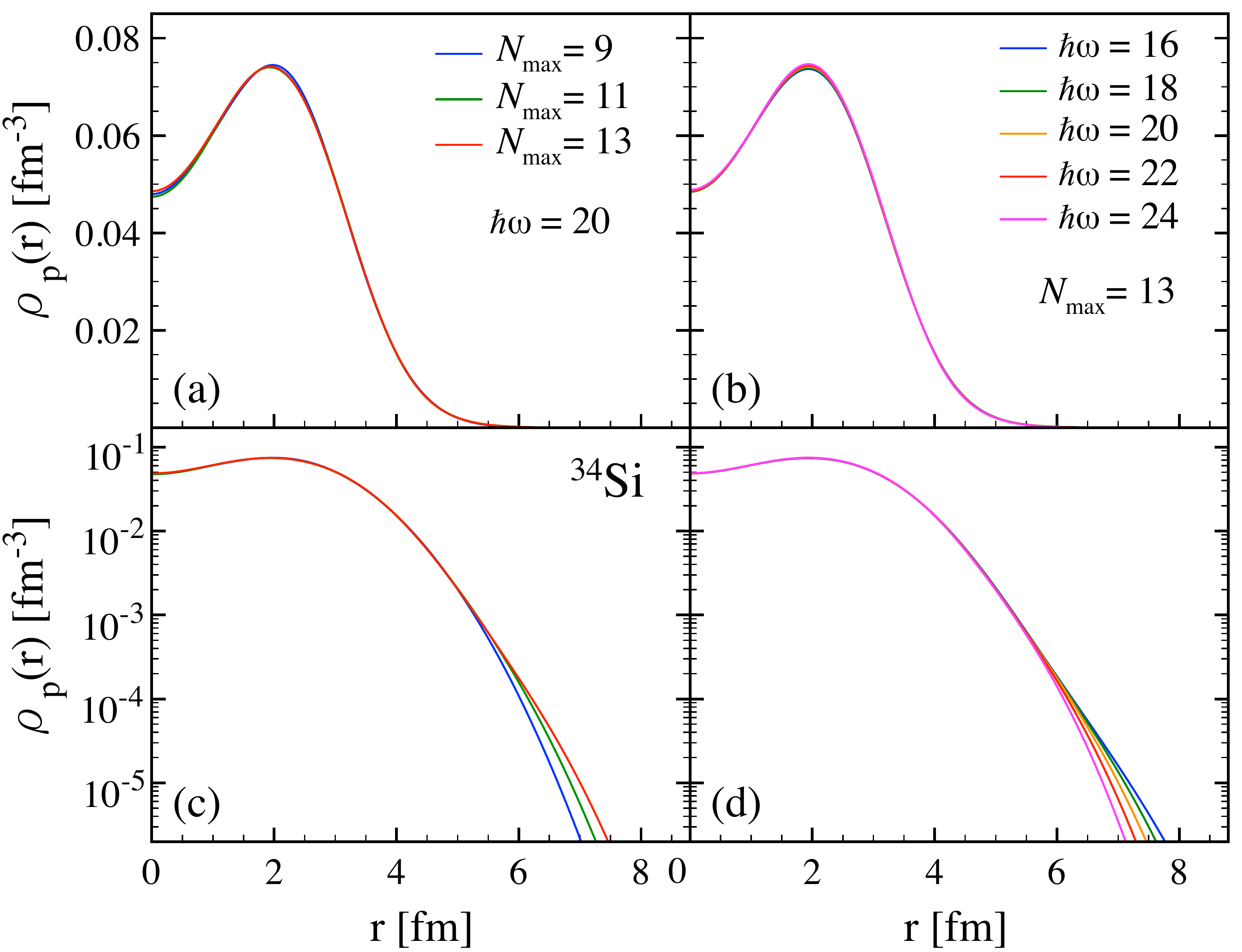}
\end{center}
\caption{(Color online) ADC(2) ground-state point-proton density distribution of $^{34}$Si for (a) different model space dimensions at $\hbar\omega=20$ MeV and (b) different harmonic oscillator frequencies at $N_{\text{{\rm max}}}=13$. Panels (c) and (d) show the same distributions respectively of (a) and (b) but with a logarithmic vertical scale.}
\label{convergenceD}
\end{figure}
Expanding the many-body Schroedinger equation on a HO basis, the spurious centre-of-mass (COM) contribution to the density distribution can be removed exactly as long as the COM part of the many-body state factorizes~\cite{Navratil:2004dp}, which is authorized by truncating the basis in term of a fixed number of A-body HO excitations. It is not presently ensured given that the basis truncation is performed directly at the level of one-body, two-body and thee-body Hilbert spaces. It happens that such a truncation procedure leads to an effective factorization of the COM part of the wave function as demonstrated in coupled-cluster calculations~\cite{Hagen:2009pq}. Although it remains to be explicitly demonstrated for SCGF calculations, the proximity of both methods and of their results gives confidence that a potential contamination of the density distribution is, at most, small in this $A=34,36$ nuclei.  More on this point in Sec.~\ref{rhoch}.

The point-proton density distribution of $^{34}$Si is displayed in Fig.~\ref{convergenceD} for different values of $N_{\text{{\rm max}}}$ at $\hbar\omega=20$ MeV and for different harmonic oscillator frequencies at $N_{\text{{\rm max}}}=13$. The overall profile shows a very weak dependence on the model space parameters. In particular, the dependence of the central density on $\hbar\omega$ is minor such that considerations about the potential bubble structure are little affected. As expected from the use of a harmonic oscillator basis, the asymptotic of the density distribution is altered by the change of $\hbar\omega$, which however does not impact the analysis presented below.

Given that we are primarily interested in features associated with potential bubble structures, we compute the point-proton $F$ parameter (Eq.~\ref{depletion_factor_bubble}) at $N_{\text{{\rm max}}}=13$ in order to estimate the uncertainty associated with the choice of $\hbar\omega$. For $\hbar\omega\in\lbrack 16,24\rbrack$\,MeV we find $F_p = 0.344~\pm~0.002$, where the error is the standard deviation of the dataset. This shows that the uncertainty associated with the choice of $\hbar\omega$ on the depletion factor is negligible compared to other sources of error (see below).

\section{Bubble structure}
\label{chap_dens}

\subsection{Point-nucleon density distributions}

The one-body density matrix $\rho$ associated with the ground state $| \Psi_0 \rangle$ of $^{34}$Si or $^{36}$S is computed in the HO basis $\{| i \rangle \equiv a^{\dagger}_i | 0 \rangle ; i\equiv n_{\text{ho}}ljm\}$ of the one-body Hilbert space ${\cal H}_1$ through 
\begin{equation}
\rho_{ij} \equiv \frac{\langle \Psi_0 | a^{\dagger}_j a_i | \Psi_0 \rangle}{\langle \Psi_0 | \Psi_0 \rangle} = \int_{C\uparrow} \frac{d\omega}{2\pi i} G_{ij}(\omega) \, , \label{density matrix}
\end{equation}
where $G_{ij}(\omega)$ denotes the (normal part of the) one-body Green's function in the energy representation~\cite{soma11a} and where the integral is performed along a closed contour located in the upper half plane of the complex plane. The point-proton density distribution reads as 
\begin{equation}
\rho_{{\rm p}}(\vec{r}) = \sum_{ij} \phi^{\ast}_{j}(\vec{r}) \phi_{i}(\vec{r})  \rho_{ij}  \, , \label{localdensity1}
\end{equation}
where $\phi_{i}(\vec{r}) \equiv \langle \vec{r} | i \rangle$ denotes HO singe-particle wave-functions and where the sum is obviously restricted to proton single-particle states. Similar expressions hold for point-neutron and matter density distributions.

\begin{figure}[t]
\centering
\includegraphics[width=8.3cm]{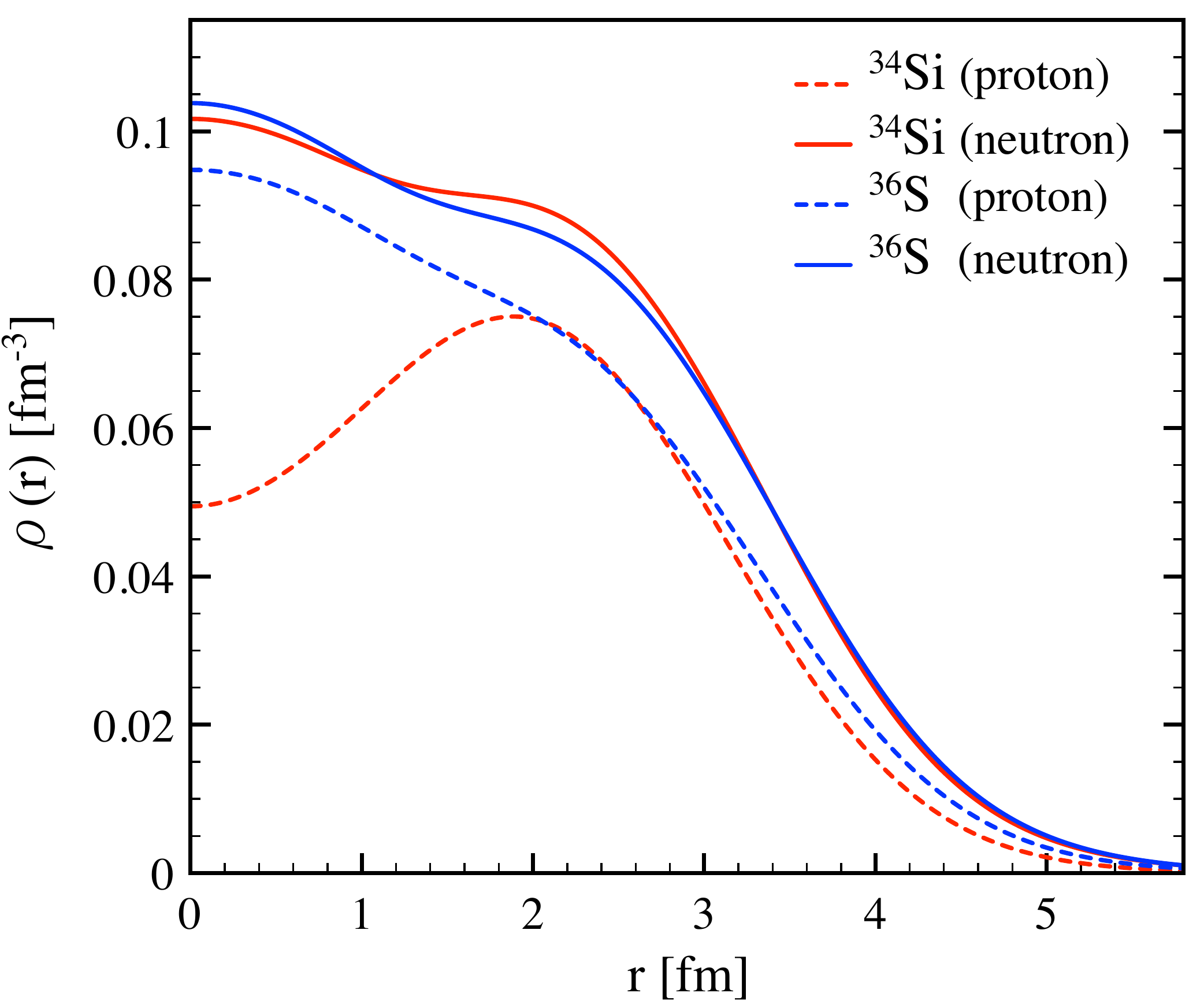}
\caption{(Color online) Point-proton and point-neutron density distributions of $^{34}$Si and $^{36}$S computed at the ADC(3) level.}
\label{fig:proton}
\end{figure}

Theoretical point-proton and point neutron density distributions of $^{34}$Si and $^{36}$S computed at the ADC(3) level are displayed in Fig.~\ref{fig:proton}. Most clearly, the point-proton density distribution of $^{34}$Si displays a pronounced depletion at the center while the one of $^{36}$S presents a maximum. This results in a noticeable (null) point-proton depletion factor $F_p=0.34$ ($F_p=0$) in $^{34}$Si ($^{36}$S). On the other hand, point-neutron density distributions are very similar, i.e. the creation of the proton bubble in $^{34}$Si associated with the removal of two protons from $^{36}$S affects the spatial distribution of neutrons only weakly by pulling them slightly away from the very center to the maximum of the proton density around $r=2$\,fm.

\subsection{Theoretical analysis}

Point-nucleon density distributions can be analyzed, internally to the theoretical scheme\footnote{See Sec.~\ref{subsec_espe} for a brief discussion on the non-observable character of quantities that are internal to the theory, i.e. that have no counterpart in the empirical world~\cite{Duguet15}, such as the presently introduced single-particle state occupations $n_{n\ell j}$~\cite{Furnstahl:2001xq}.}, by expressing them in the so-called natural basis $\{| \mu \rangle \equiv b^{\dagger}_\mu | 0 \rangle ; \mu\equiv n\ell jm\}$, i.e. in the orthonormal basis of ${\cal H}_1$ that diagonalizes the one-body density matrix $\rho$. Natural orbitals are thus obtained in the HO basis by solving the eigenvalue equation
\begin{equation}
\sum_{j} \rho_{ij} \langle j | \mu \rangle = n_{\mu} \langle i | \mu \rangle  \, . \label{diago_rho}
\end{equation}
In the natural basis, the point-proton density distribution reduces to a sum of positive single-particle contributions. Taking into account the spherical symmetry characterizing the $J^{\pi}=0^+$ ground state of the nuclei of interest, the point-proton density distribution eventually reads as
\begin{equation}
\rho_{{\rm p}}(\vec{r}) = \sum_{n\ell j} \frac{2j+1}{4\pi} n_{n\ell j} R^{2}_{n\ell j}(r) \equiv \sum_{\ell j} \rho_{{\rm p}}^{\ell j}(r)\, , \label{dens_natural}
\end{equation}
where $R_{n\ell j}(r)$ denotes the radial part of the natural single-particle wave-functions $\varphi_{n\ell jm}(\vec{r})$ and where the partial-wave contributions $(\ell ,j)$ to the density have also been introduced.

\begin{figure}[t]
\centering
\includegraphics[width=8.3cm]{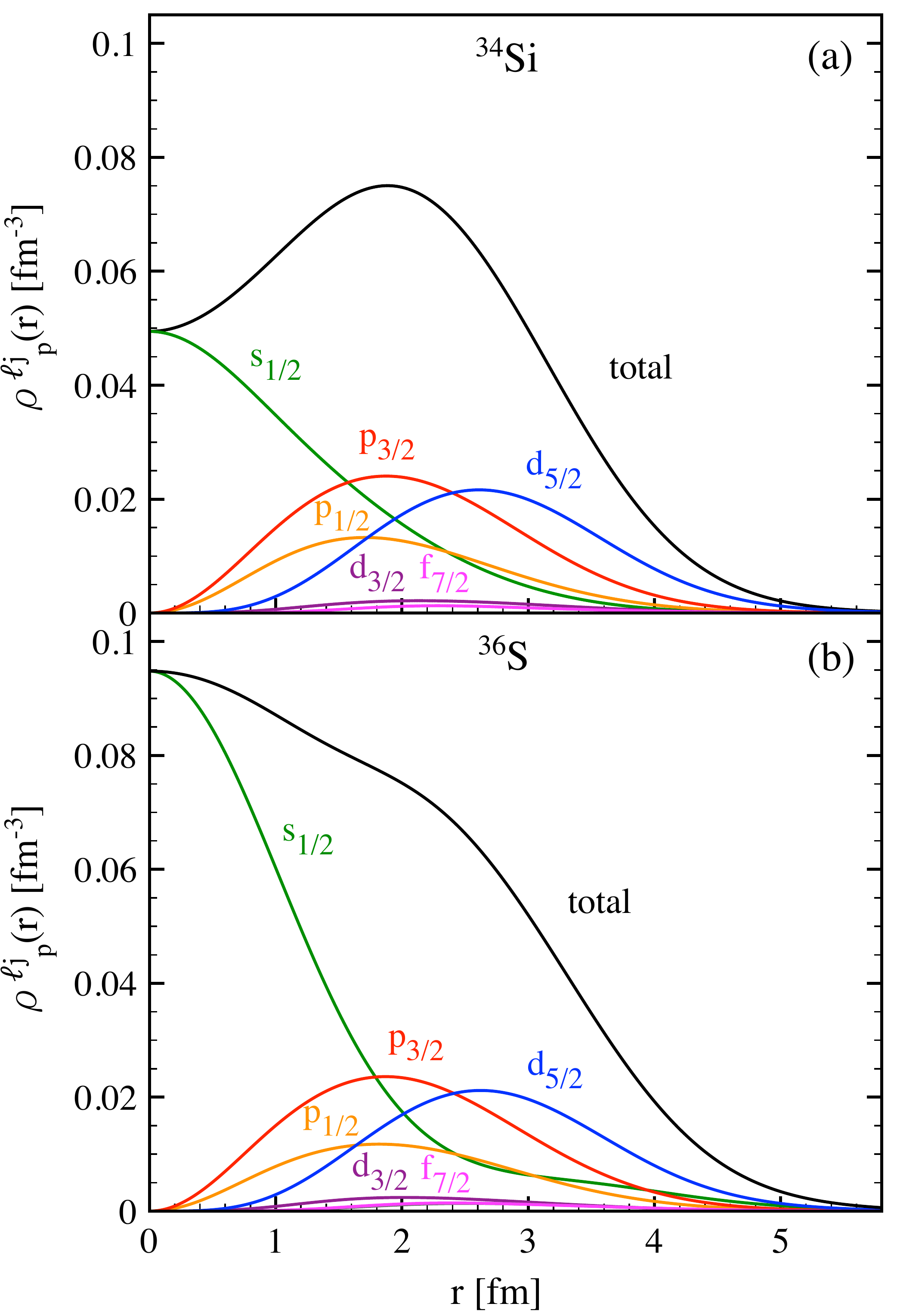}
\caption{(Color online) Natural orbital partial-wave decomposition $\rho_{{\rm p}}^{\ell j}(r)$ of the point-proton density distribution computed at the ADC(3) level for $^{34}$Si (a) and $^{36}$S (b).}
\label{decomposition_density34Si}
\end{figure}

Figure~\ref{decomposition_density34Si} displays the partial-wave decomposition of point-proton density distributions of $^{34}$Si and $^{36}$S. In $^{34}$Si, the very interior of the density is entirely built from the $s_{1/2}$ partial-wave and is depleted compared to its maximum at about $r=1.9$\,fm that is dictated by the $p_{3/2}$ and $p_{1/2}$ partial-waves. The $d_{5/2}$ wave contributes at larger radii and dominates at the surface. One also observes small contributions from the $d_{3/2}$ and $f_{7/2}$ waves.

\begin{figure}[t]
\centering
\includegraphics[width=7.5cm]{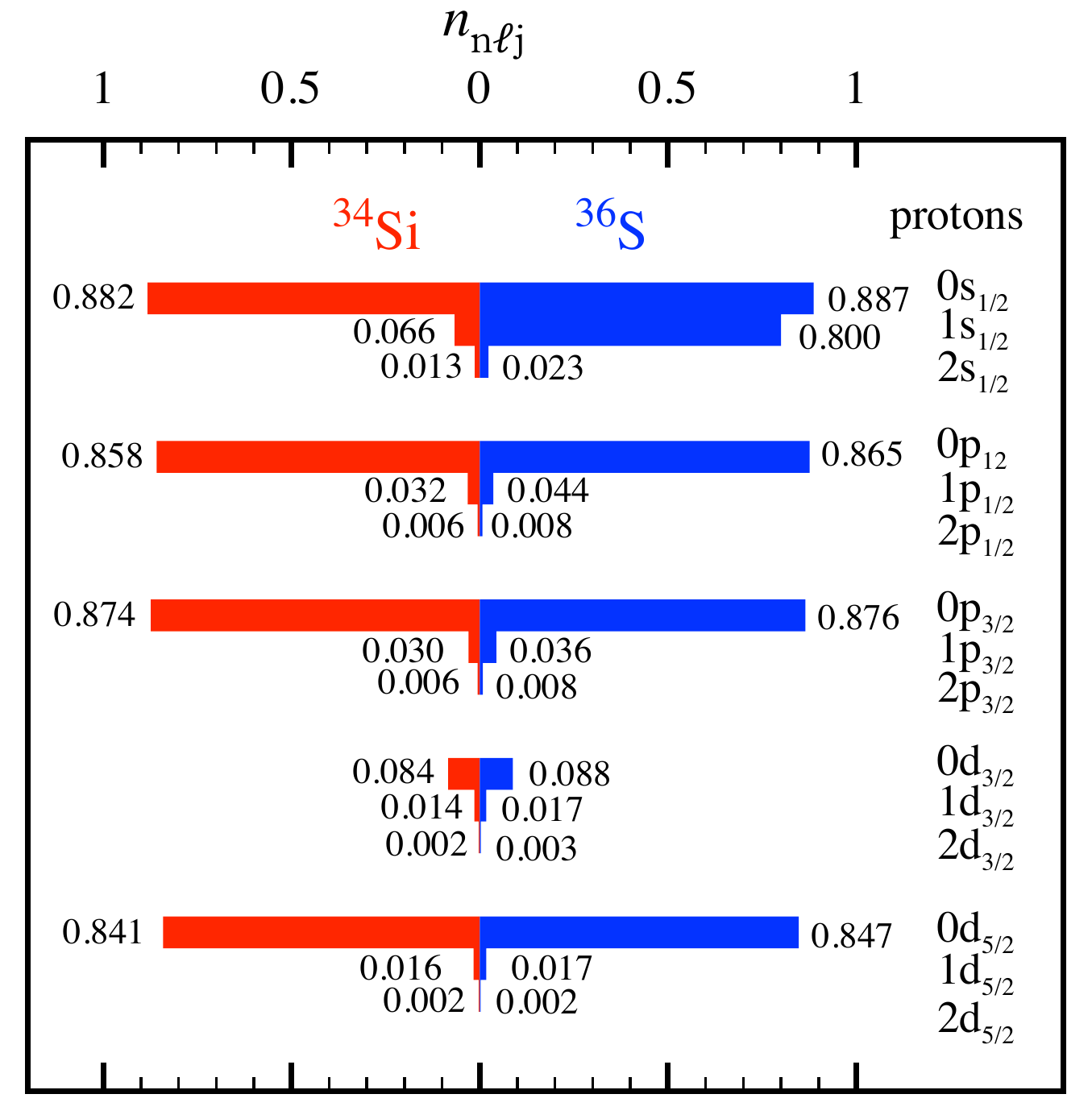}
\caption{(Color online) Proton natural orbitals occupations $n_{n\ell j}$ computed at the ADC(3) level in $^{34}$Si and $^{36}$S.}
\label{prot_occ_density}
\end{figure}

\begin{figure}[t]
\centering
\includegraphics[width=8cm]{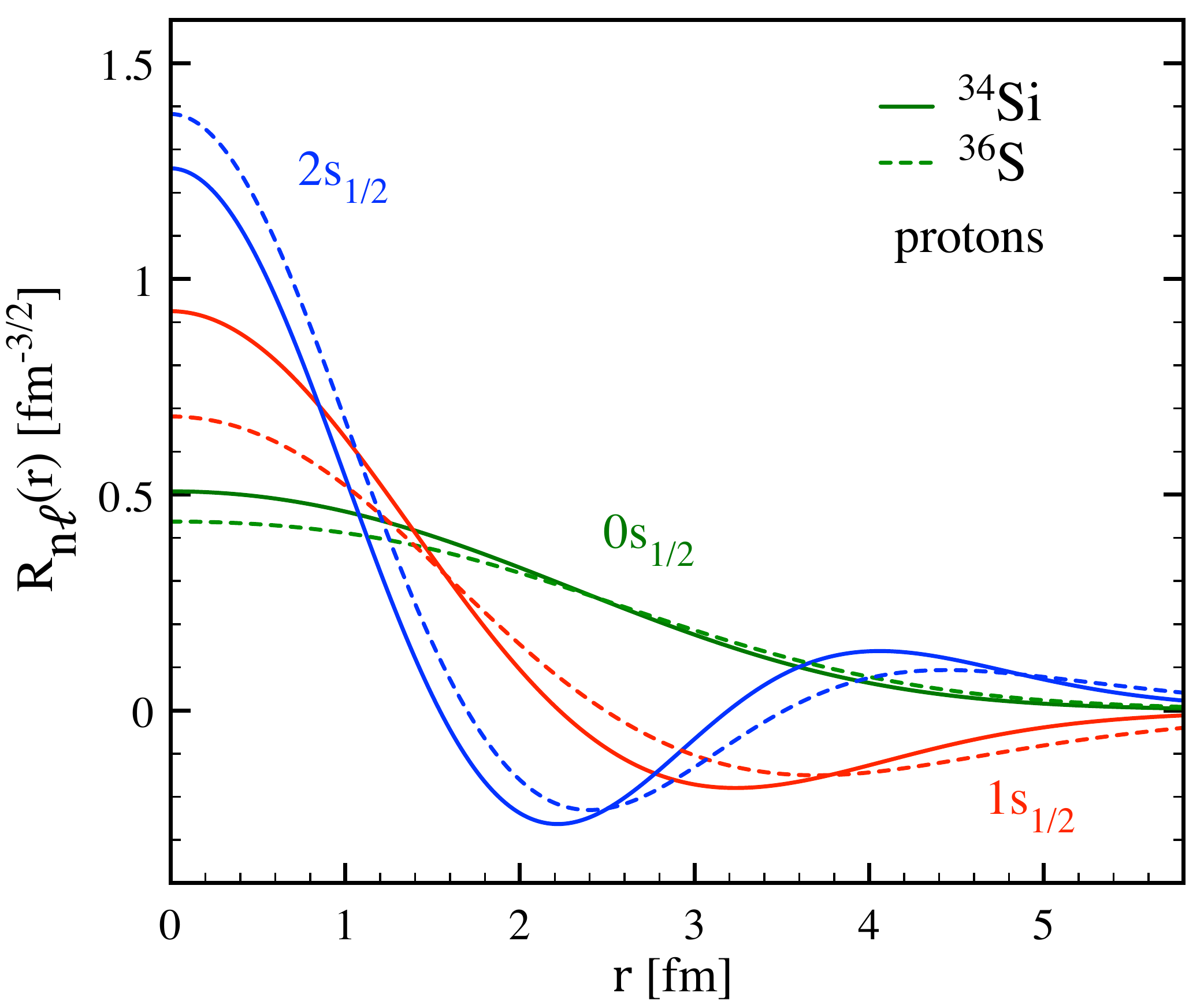}
\caption{(Color online) Radial part of proton $0s_{1/2}$, $1s_{1/2}$ and $2s_{1/2}$ natural single-particle wave-functions calculated at the ADC(3) level in $^{34}$Si and $^{36}$S.}
\label{wfs_density34Si36S}
\end{figure}

Ab initio calculations describe the complete dynamics of the $A$ interacting nucleons in large model spaces. Consequently, each $(\ell ,j)$ partial wave builds from several orbitals corresponding to different principal quantum numbers $n$. Figure~\ref{prot_occ_density} decomposes the occupation of each partial wave into individual proton natural orbital occupation. As expected, the $s_{1/2}$ partial-wave is dominated by the $1.8$ protons occupying the $0s_{1/2}$ states. Still, the $0.13$ ($0.03$) protons occupying the $1s_{1/2}$ ($2s_{1/2}$) states do contribute for $18\%$ ($7\%$) of the density at $r=0$. This surprisingly large contribution originates from the fact that the $1s_{1/2}$ ($2s_{1/2}$) wave-function is $1.8$ ($2.4$) times larger than the $0s_{1/2}$ wave-function at $r=0$ (see Fig.~\ref{wfs_density34Si36S}). 

\begin{figure}[t]
\centering
\includegraphics[width=7.5cm]{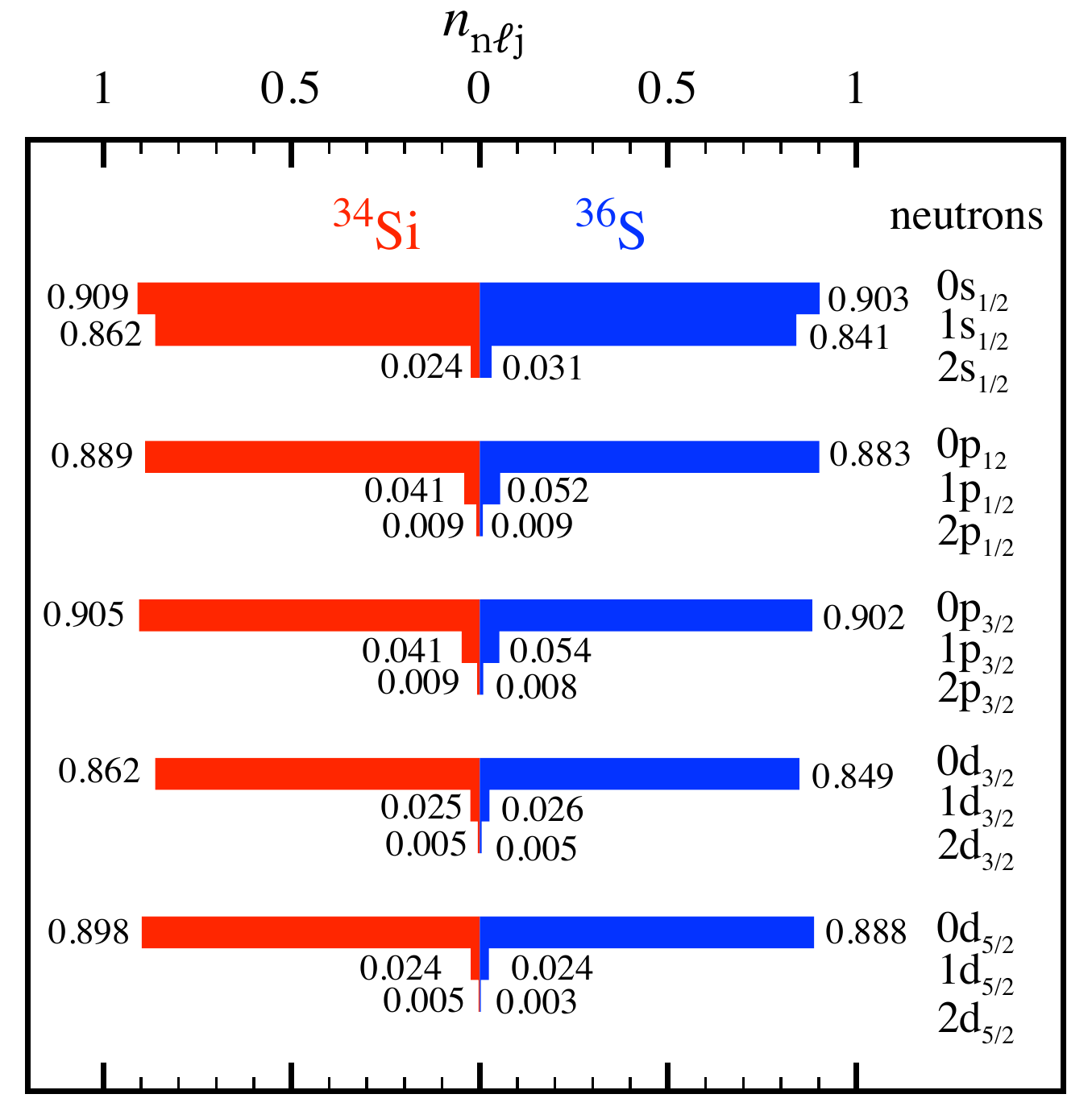}
\caption{(Color online) Same as Fig.~\ref{prot_occ_density} for neutrons.}
\label{neutr_occ_density}
\end{figure}

As visible in Fig.~\ref{decomposition_density34Si}, the change in point-proton density from $^{34}$Si to $^{36}$S entirely originates from the $s_{1/2}$ partial wave. The proton density almost doubles at $r=0$ when adding two protons to $^{34}$Si, leading to the disappearance of the bubble structure in $^{36}$S. As expected from a naive independent particle picture, the rise of the central density is driven by the $1s_{1/2}$ states whose occupation increases from $0.1$ in $^{34}$Si to $1.6$ protons in $^{36}$S. Interestingly though, this increase in occupation is  somewhat compensated by a lowering of the associated wave function at $r=0$, as is visible from Fig.~\ref{wfs_density34Si36S}. Consequently, the rise of the central density of $^{36}$S due to the increase of $n_{1s_{1/2}}$ is only 55$\%$ of what it would have been with a frozen $1s_{1/2}$ natural orbital wave-function. Correspondingly, the $0s_{1/2}$ ($2s_{1/2}$) states decrease (increase) the central density by an amount that corresponds to 20$\%$ (8$\%$) of the rise generated by the $1s_{1/2}$ states. These subleading contributions mainly originate from the fact that the two added protons lead to a lowering (rising) of the $0s_{1/2}$ ($2s_{1/2}$) natural wave-functions at $r=0$. Overall, even though the disappearance of the proton bubble when going from $^{34}$Si to $^{36}$S does mainly reflect the increased occupation of the $1s_{1/2}$ states, one observes that in quantitative terms the net effect results from the combination of several intricate features in our ab initio calculations.

As testified by Fig.~\ref{neutr_occ_density}, the fact that point-neutron density distributions of $^{34}$Si and $^{36}$S are (nearly) identical reflects the (essentially) equal occupations of neutron natural orbitals and their (essentially) unchanged wave functions.

\subsection{Impact of correlations}

\begin{figure}[b]
\begin{center}
\centering
\includegraphics[width=8.6cm]{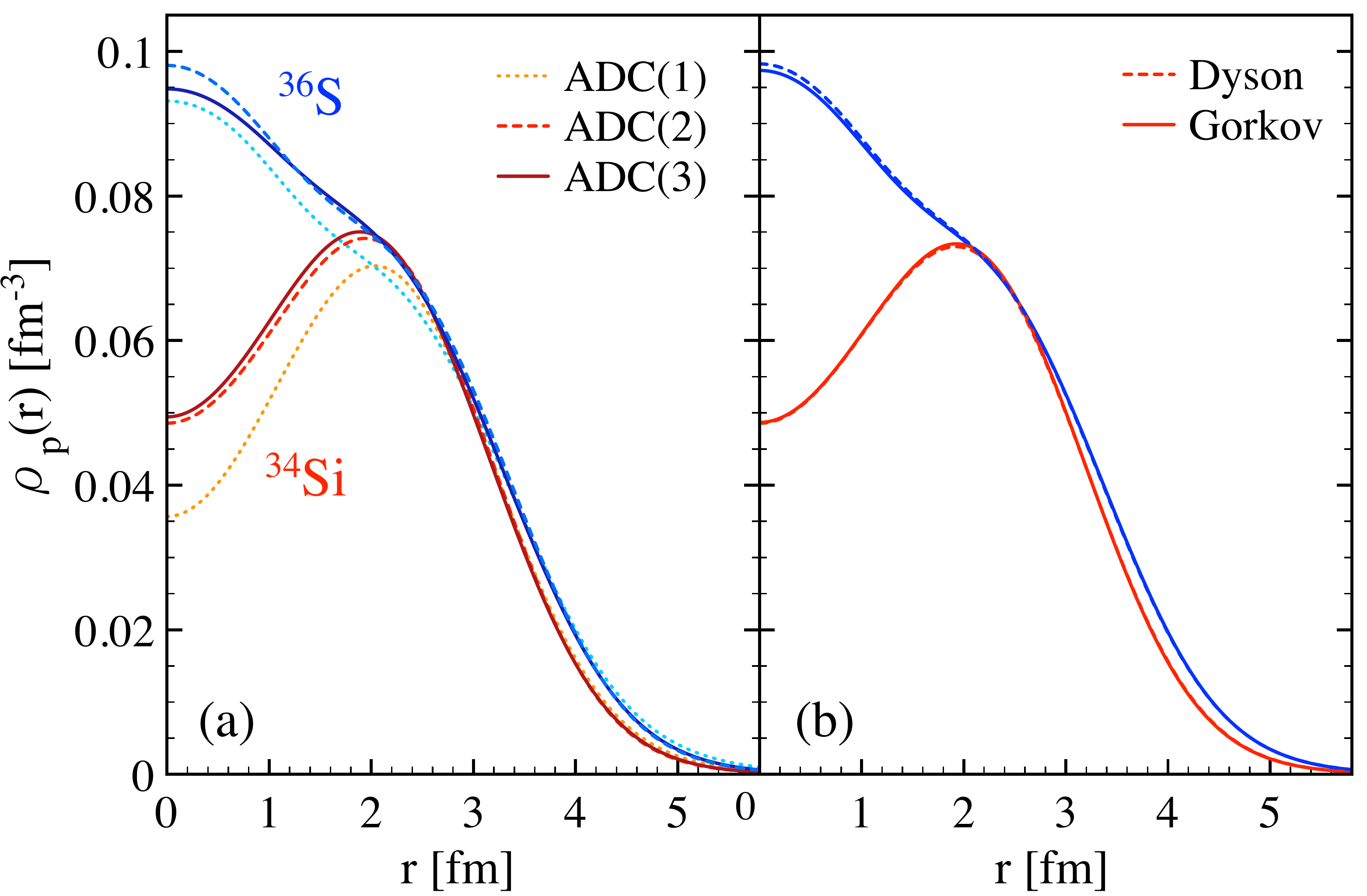}
\end{center}
\caption{(Color online) Point-proton density distributions of $^{34}$Si and $^{36}$S. (a) Results from Dyson ADC(1), ADC(2) and ADC(3) calculations. (b) Results from Gorkov and Dyson ADC(2) calculations.}
\label{fig:impact_correlations}
\end{figure}

In panel (a) of Fig.~\ref{fig:impact_correlations}, point-proton density distributions of $^{34}$Si and $^{36}$S are compared at various levels of many-body truncations, i.e. as obtained from Dyson ADC(1), ADC(2) and ADC(3) calculations. Moving from ADC(1) (i.e. HF) to ADC(2), the amplitude of the central depletion diminishes in $^{34}$Si. This erosion of the bubble structure is the consequence of explicit dynamical correlations added at the ADC(2) level, knowing that correlations added at the ADC(3) level do not further change the picture. Eventually, this erosion results in a decrease of the point-proton F factor from $0.49$ to $0.34$ when going from HF to ADC(3) calculations (see Tab.~\ref{correlationsF}). The impact of correlations on the point-proton density distribution of $^{36}$S is less pronounced.

\begin{table}
\centering
\begin{tabular}{|c||c|c|c|}
\hline
$^{34}$Si & ADC(1) & ADC(2) & ADC(3) \\
\hline
\hline
$F_p$ & 0.49 & 0.34 & 0.34 \\
\hline
\end{tabular}
\caption{Point-proton depletion factor in $^{34}$Si computed within ADC(1), ADC(2) and ADC(3) approximations.}
\label{correlationsF}
\end{table}

The impact of many-body correlations can be analyzed on the basis of the natural orbital decomposition of the density. Given that the one-body density matrix $\rho$ reflects the correlations included in the calculation of $| \Psi_0 \rangle$, it is clear that not only the natural occupations but also the natural orbital wave-functions change with the many-body truncation scheme employed.

\begin{figure}[t]
\centering
\includegraphics[width=8cm]{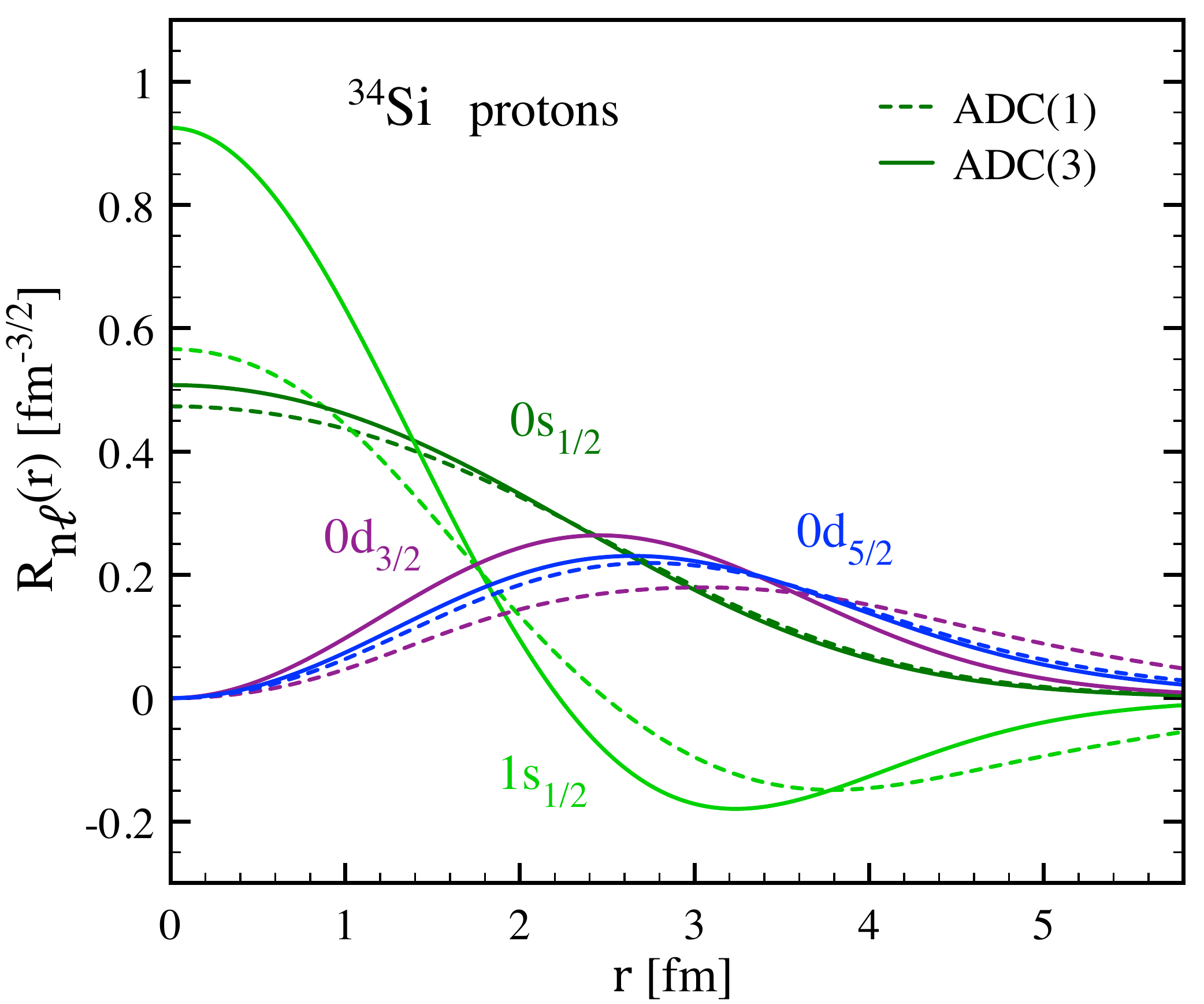}
\caption{(Color online) Radial part of selected natural single-particle wave-functions $\varphi_{n\ell jm}(\vec{r})$ in $^{34}$Si. Results are displayed at the ADC(1) and ADC(3) level.}
\label{wfs_density34Si}
\end{figure}

\begin{figure}[t]
\centering
\includegraphics[width=7.5cm]{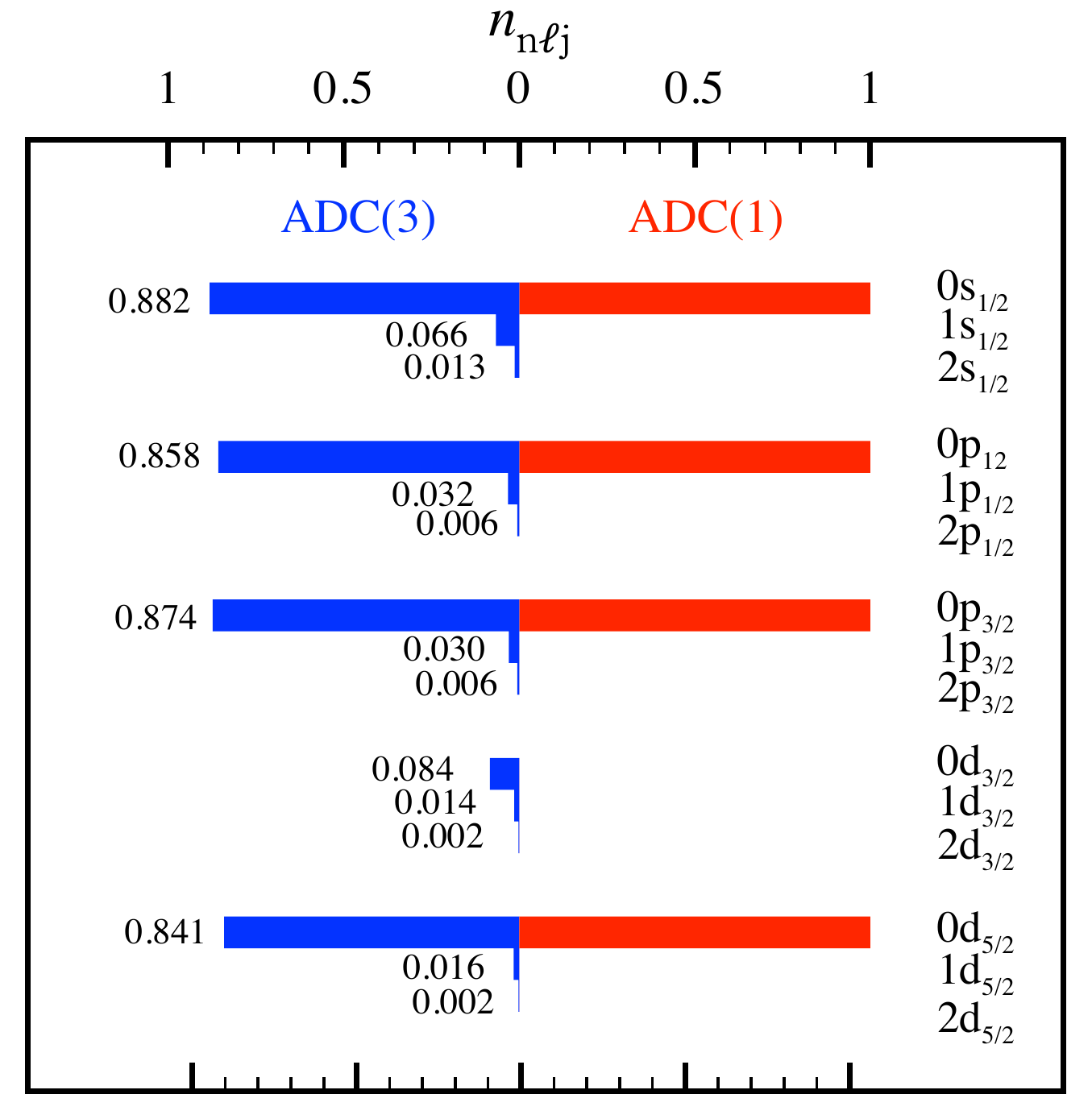}
\caption{(Color online) Natural orbitals occupations $n_{n\ell j}$ in $^{34}$Si. Results are displayed at the ADC(1) and ADC(3) levels.}
\label{occ_density34Si}
\end{figure}

First, dynamical correlations partially promote protons from  ($0s_{1/2}$, $0p_{3/2}$, $0p_{1/2}$, $0d_{5/2}$) states into ($1s_{1/2}$, $1p_{3/2}$, $1p_{1/2}$, $0d_{3/2}$, $1d_{5/2}$) as is visible from Fig.~\ref{occ_density34Si}. Second, long-range correlations lead to a contraction of the proton natural wave-functions, the effect being the most drastic for the orbitals that are originally unoccupied at the ADC(1) level, e.g. for the  $1s_{1/2}$ and $0d_{3/2}$ wave-functions on Fig.~\ref{wfs_density34Si}. This is particularly true for the $1s_{1/2}$ wave-function that becomes strongly peaked at $r=0$. In spite of partially promoting protons into orbitals that are located more towards the outside, long-range correlations induce a reduction of the charge rms radius of the nucleus associated with the contraction of the natural orbital wave functions. With the presently used NNLO$_{\text{sat}}$ Hamiltonian, this further improves the agreement of the predicted charge rms radius of $^{36}$S with experimental data as visible in Tab.~\ref{radiicorrelations}.

\begin{figure}[t]
\centering
\includegraphics[width=8.3cm]{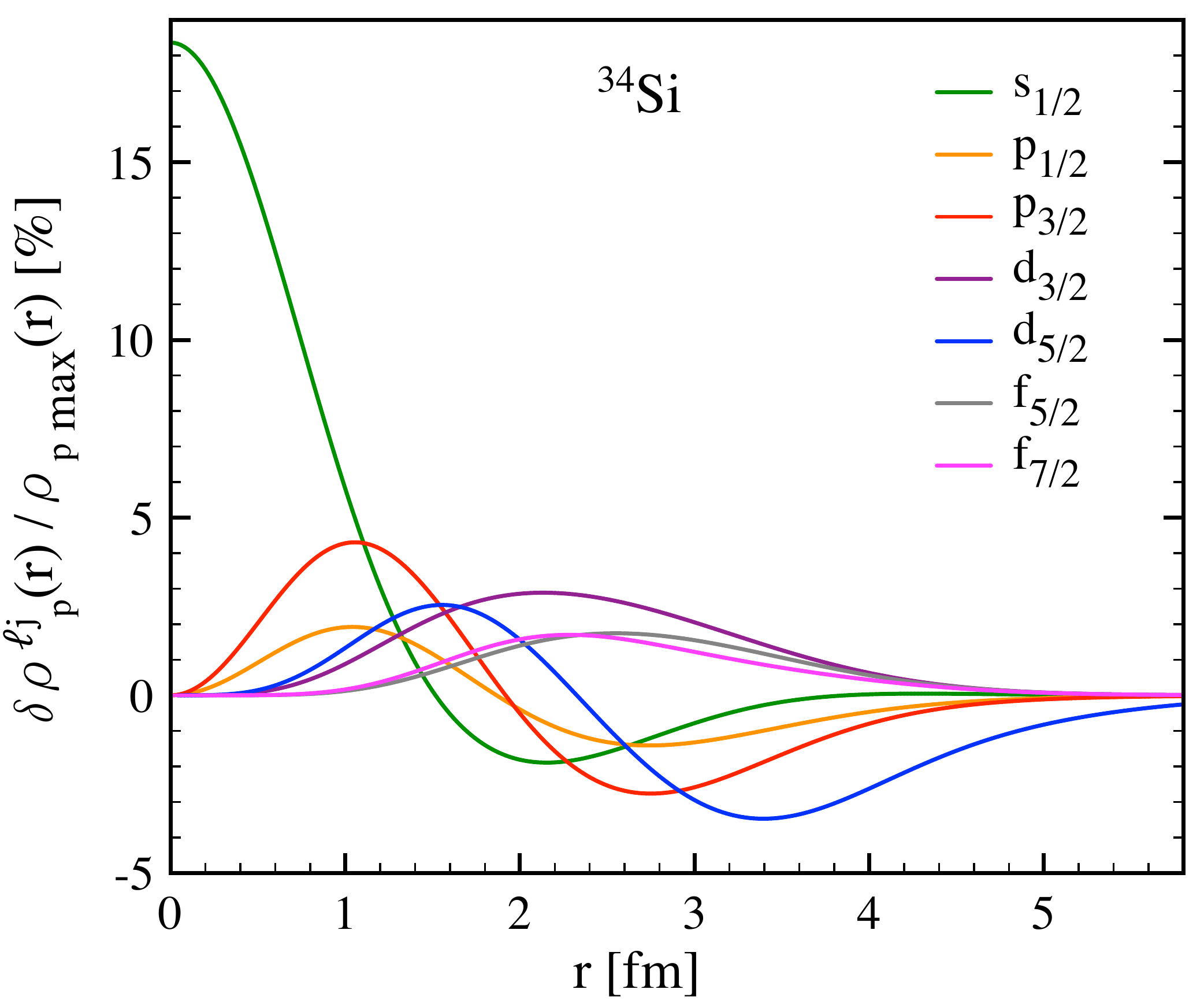}
\caption{(Color online) Partial-wave contributions $\delta \rho^{\ell j}_{{\rm p}}(r)$ to the difference between point-proton density distributions of $^{34}$Si computed at the ADC(3) and ADC(1) levels.}
\label{var_dens_density34Si}
\end{figure}

The net result of many-body correlations on the point-proton density of $^{34}$Si is visible in Fig.~\ref{var_dens_density34Si}, where the variation due to each partial wave is displayed. One observes that the non-zero occupation of the $1s_{1/2}$ states and the contraction of their wave function increases the central density by about 18$\%$ of its maximum value, leading to the erosion of the bubble structure mentioned above.  The fact that the bubble structure can be significantly suppressed by the inclusion of long-range correlations was already pointed out on the basis of MR-EDF~\cite{Yao12,Yao13} and shell-model (SM)~\cite{Grasso09} calculations.

On panel (b) of Fig.~\ref{fig:impact_correlations}, point-proton density distributions of $^{34}$Si and $^{36}$S obtained from both Dyson and Gorkov SCGF calculations at the ADC(2) level are compared. Pairing correlations are very weak in these two closed sub-shell nuclei such that their explicit account does not impact the results in any significant way. This feature, presently obtained on the basis of realistic 2N and 3N inter-nucleon interaction, mirrors the situation at play in SR-EDF calculations~\cite{Grasso09}.

\subsection{Charge density distribution}
\label{rhoch}

Generically speaking, the electromagnetic charge density (and current) operator is expressed as an expansion in many-body operators acting on nucleonic degrees of freedom. This operator not only accounts for the point distribution of protons but also for their own charge distribution, along with the one of neutrons, and for charge (and current) distributions associated with the light charged mesons they exchange. 
To first approximation, the nuclear charge density can be obtained through the folding of the nuclear point-proton density distribution with the charge density distribution of the proton. In doing so, one omits neutrons' contribution\footnote{We remind however that neutron's charge density contribution to \textit{charge radii} is presently taken into account~\cite{Cipollone15}.} as well as relativistic spin-orbit corrections, both typically relatively small~\cite{chandra76a}.
We thus compute the charge density distribution as
\begin{equation}\label{charge_density}
\rho_{{\rm ch}}(r) = \hspace{-.05cm}  \sum_{i=1}^{3} \frac{\theta_i}{r_i\sqrt{\pi}} \hspace{-.1cm} \int\limits_{0}^{+\infty} \hspace{-.1cm}  dr' \frac{r'}{r} \rho_{{\rm p}}(r')\left\lbrack \text{e}^{-\left(\frac{r-r'}{r_i}\right)^{2}} \hspace{-.15cm}  - \text{e}^{-\left(\frac{r+r'}{r_i}\right)^{2}} \right\rbrack ,
\end{equation}
where the sets $(\theta_i,r_i)$ come from having parameterized the charge density distribution of the proton as a linear superposition of three Gaussians and have been adjusted to reproduce the proton charge form factor from electron scattering data~\cite{chandra76a}.
The proton rms radius that results from this parameterization is $\langle R^2_p \rangle^{1/2} =~0.88~\text{fm}$, consistent with the value used to compute the rms charge radius~\cite{CODATA2012}. Let us note that eventual smaller values of $\langle R^2_p \rangle^{1/2}$~\cite{Pohl2010} would lead to an increase of the depletion factor (see also Tab.~\ref{F34SiEDF} and corresponding discussion).

Furthermore, one needs to correct for spurious center of mass and include Darwin-Foldy relativistic correction. Assuming\footnote{While this has been proven for coupled-cluster~\cite{Hagen:2009pq} and in-medium similarity renormalization group~\cite{hergert16a} calculations, a similar study remains to be done for SCGF calculations. Given the proximity of these many-body methods, one is however confident that the center of mass factorization does indeed occur in the same way in SCGF calculations as is assumed here.} that the center of mass wave-function factorizes in the ground-state of a harmonic oscillator Hamiltonian characterized by the frequency $\tilde{\omega}$, the inclusion of spurious center-of-mass and Darwin-Foldy relativistic corrections can be performed at the price of proceeding to the replacement~\cite{negele70, chandra76a}
\begin{equation} 
r^2_i \longrightarrow r^2_i  - \frac{b^2}{A} + \frac{1}{2}\left(\frac{\hbar}{m}\right)^2
\end{equation}
in Eq.~\ref{charge_density},  where $m$ is the nucleon mass, hence~\footnote{We use here, as everywhere throughout the paper, $c=1$.} $\hbar/m= 0.21$\,fm, and $b^2=(m \, \hbar \, \tilde{\omega})^{-1}$.
Employing Bethe's formula~\cite{negele70}, the latter term can be approximated with $b^2 \approx A^{1/3} \,\,\text{fm}^2$. We note that, for $^{16}$O, such an approximation is consistent with the value of $\hbar\tilde{\omega}$ found in Ref.~\cite{Hagen:2009pq} and is thus safe to use in present calculations of $^{34}$Si and $^{36}$S.

\begin{figure}[t]
\centering
\includegraphics[width=8.3cm]{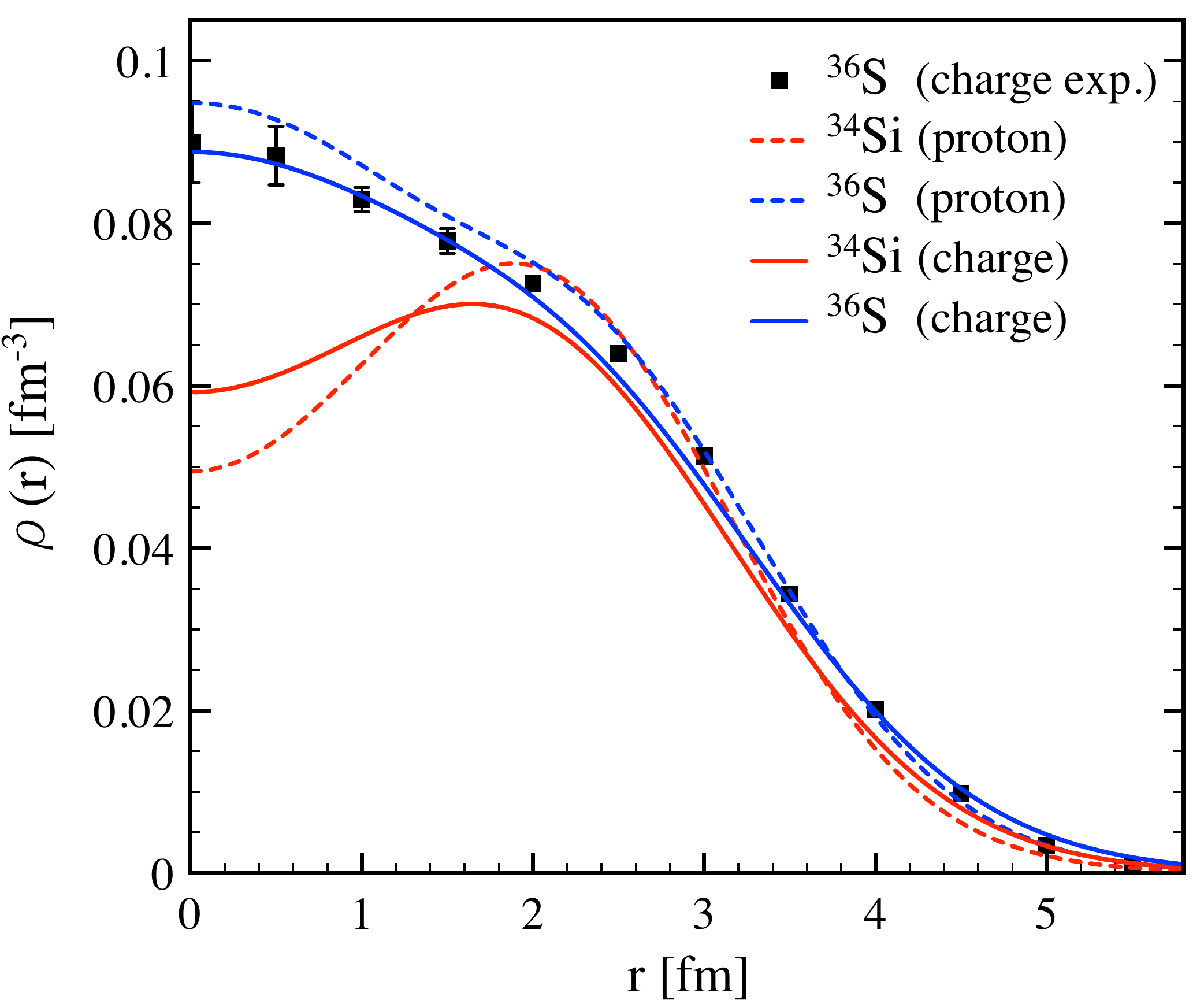}
\caption{(Color online) Charge and proton densities of $^{34}$Si and $^{36}$S at the ADC(3) level. The experimental charge density of $^{36}$S is taken from Ref.~\cite{Rychel83}.}
\label{fig:charge_proton}
\end{figure}

Theoretical charge density distributions of $^{34}$Si and $^{36}$S computed at the ADC(3) level are compared to their point-proton counterpart in Fig.~\ref{fig:charge_proton} and to the experimental charge density of $^{36}$S~\cite{Rychel83}. The excellent agreement between theoretical and experimental charge density distributions of $^{36}$S gives confidence in the SCGF prediction obtained with NNLO$_{\text{sat}}$ for $^{34}$Si. While the folding operated in Eq.~\ref{charge_density} weakly reduces the peak at the center of the density distribution of $^{36}$S, it significantly smears out the depletion in the point-proton density distribution of $^{34}$Si. This effect could be expected given that the folding takes place over a distance set by the proton charge radius that is consistent with the size of the proton bubble. The fact that the bubble structure could be strongly suppressed in the observable charge density of $^{34}$Si was already pointed out on the basis of SR- and MR-EDF calculations~\cite{Yao12,Yao13}. This reflects in the value of the F factor of $^{34}$Si that goes down from $0.34$ to $0.15$ when going from the point-proton to the charge density distribution (see Tab.~\ref{F34SiEDF}).

\subsection{Form factor}

Accessing the charge density distribution of $^{34}$Si would require to scatter electrons on radioactive ions. This would lead to measuring the electromagnetic charge form factor, which relates to the nuclear charge density distribution through
\begin{equation}
F(q) = \int d\vec{r} \rho_{{\rm ch}}(r)e^{-i\vec{q} \cdot \vec{r}}\, ,
\end{equation} 
where $\vec{q}$ is the transferred momentum, itself related to the incident momentum $\vec{p}$ and the scattering angle $\theta$ via $q = 2p\sin\theta/2$. 

\begin{figure}[b]
\centering
\includegraphics[width=8.6cm]{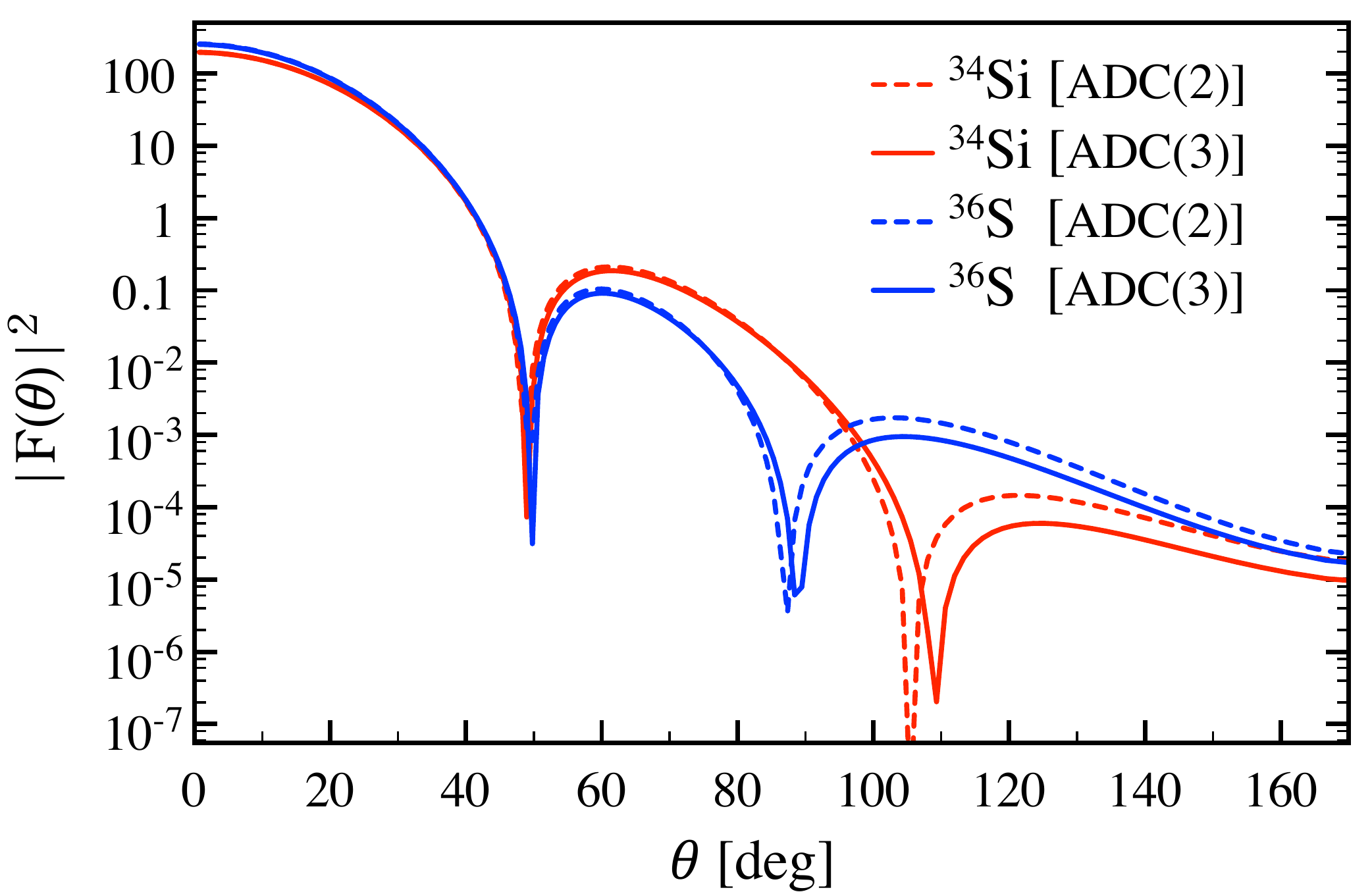}
\caption{(Color online) Angular dependence of the form factor obtained for 300 MeV electron scattering on $^{34}$Si and $^{36}$S. Results from both ADC(2) and ADC(3) calculations are displayed.}
\label{fig:form_factor}
\end{figure}

In Ref.~\cite{Khan07}, Hartree-Fock densities based on two Skyrme EDF parameterizations were used to demonstrate that the diffraction pattern of a semi-bubble nucleus differs significantly from the one of the same nucleus without a bubble. Similarly, Fig.~\ref{fig:form_factor} displays the angular dependence of the form factor obtained at the ADC(2) and ADC(3) levels for 300 MeV electron scattering on $^{34}$Si and $^{36}$S. From 60 to 95 degrees, the angular distribution is located at higher magnitude in $^{34}$Si than in $^{36}$S. This is accompanied by a shift of about 20 degrees between both second minima such that the two angular distributions are out of phase at about 110 degrees. Furthermore, the comparison between ADC(2) and ADC(3) results demonstrates that this pattern is solid against the further addition of many-body correlations whenever the bulk of them have been included. All in all, the depletion in the charge density of  $^{34}$Si is predicted to be large enough for the fingerprint of the bubble structure to be potentially visible in a future electron scattering experiment. 

\subsection{Choice of the Hamiltonian}
\label{choiceH}

\begin{figure}[b]
\begin{center}
\centering
\includegraphics[width=8.6cm]{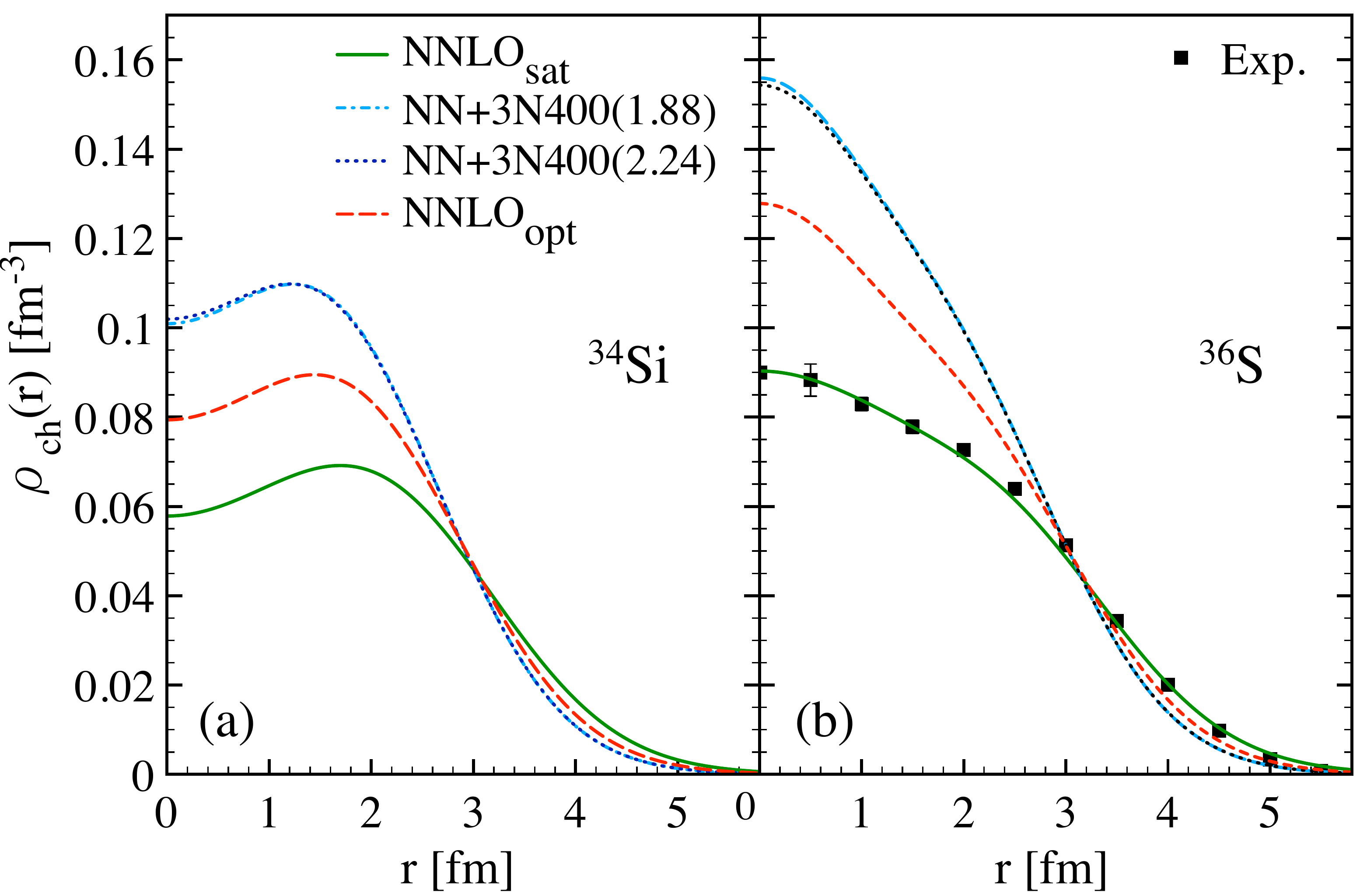}
\end{center}
\caption{(Color online) Charge density distributions computed at the ADC(2) level with four different (2N+3N) interactions for $^{34}$Si (a) and $^{36}$S (b). The experimental charge density of $^{36}$S is also shown~\cite{Rychel83}.}
\label{hamiltonian}
\end{figure}

We wish to probe the sensitivity of the results to the choice of the input nuclear Hamiltonian. Ideally, one would like to propagate systematic uncertainties associated with the $\chi$EFT-based Hamiltonian to many-body observables of interest\footnote{This should be performed at various chiral orders, i.e. going from LO to NLO, to NNLO etc, in order to check the consistency of the extrapolated uncertainties.}. Although efforts in that direction are currently being made~\cite{Carlsson:2015vda}, it is not possible to implement this program convincingly yet. Consequently, we presently limit ourselves to repeating the calculation for the four representative $\chi$EFT-based Hamiltonians introduced in Sec.~\ref{calcsetup}.

\begin{table}
\centering
\begin{tabular}{|c||c|c|c|}
\hline
$^{36}$S & NN+3N400(1.88) & NNLO$_{\text{opt}}$ & NNLO$_{\text{sat}}$  \\
\hline
\hline
$\langle r^{2}_{{\rm ch}}\rangle^{1/2}$ & 2.864 & 3.033 & 3.291  \\
\hline
\end{tabular}
\caption{Charge rms radii (in fm) of $^{36}$S computed at the ADC(2) level from four different chiral-EFT(-based) Hamiltonians. Experimentally~\cite{Angeli13}, $\langle r^{2}_{{\rm ch}}\rangle^{1/2}$ = 3.2985 $\pm$ 0.0024 fm.}
\label{radiihamiltonian}
\end{table}

The charge density distribution of $^{34}$Si ($^{36}$S) computed at the ADC(2) level from the four different Hamiltonians is displayed in the left (right) panel of Fig.~\ref{hamiltonian}. Starting with $^{36}$S for which experimental data exist, one can appreciate the significant spread of the results. While the charge density obtained with NNLO$_{\text{sat}}$ is in close agreement with data, those obtained from NNLO$_{\text{opt}}$, NN+3N400(1.88) and NN+3N400(2.24) are much too large in the bulk and do not extend far enough. The shortcoming of the last three Hamiltonians relates to their incapacity to account for both nuclear binding and nuclear size at the same time~\cite{Soma:2013xha, Hergert:2014iaa,Lapoux:2016exf}. As visible in Tab.~\ref{radiihamiltonian}, the pattern in the charge density does correlate with the charge rms radius. As a matter of fact, fixing the latter by using charge radii of $^{14}$C and $^{16}$O in the fit of NNLO$_{\text{sat}}$, the overall charge density is very well reproduced. Empirically speaking, and omitting the questions raised by this way of fixing the problem, this result strongly favors the prediction made with NNLO$_{\text{sat}}$ in the present study.

The variability of the charge density of $^{34}$Si follows the same pattern as in $^{36}$S and thus reflects the capacity of the Hamiltonian to account for nuclear sizes. The choice of the Hamiltonian slightly impacts the radial extension of the bubble but not its depth on an absolute scale that appears to be a robust prediction. However, as the density in the bulk overall shrinks as one goes from NN+3N400, to NNLO$_{\text{opt}}$ and to NNLO$_{\text{sat}}$, the F factor grows accordingly as reported in Tab.~\ref{interactionsF}. Given the empirical superiority of NNLO$_{\text{sat}}$, the corresponding value ($F_{{\rm ch}}=0.15$ at both ADC(2) and ADC(3) levels) provides the most probable ab initio characterization of the charge bubble in $^{34}$Si as of today.

Last but not least, one notices that varying the momentum scale of the SRG transformation applied to the NN+3N400 Hamiltonian from $2.24$\,fm$^{-1}$ to $1.88$\,fm$^{-1}$ has negligible impact on the charge density of $^{34}$Si and $^{36}$S. However, it is to be noticed that comparing results obtained from the NN+3N400 Hamiltonian to which a SRG transformation was applied to those obtained from the unevolved NNLO$_{\text{sat}}$ Hamiltonian raises the question of the SRG transformation of the charge density operator in the former case. Although such an evolution is unlikely to impact the bubble structure significantly, it is a caveat to be mentioned here and investigated in the future.
\begin{table}
\centering
\begin{tabular}{|c||c|c|c|}
\hline
$^{34}$Si & NN+3N400(1.88) & NNLO$_{\text{opt}}$ & NNLO$_{\text{sat}}$ \\
\hline
\hline
$F_{{\rm ch}}$ & 0.08 & 0.11  & 0.15 \\
\hline
\end{tabular}
\caption{Charge depletion factors in $^{34}$Si computed with various $\chi$EFT interactions at the ADC(2) level.}
\label{interactionsF}
\end{table}

\subsection{Impact of 3N interactions}

\begin{figure}[b]
\centering
\includegraphics[width=8.6cm]{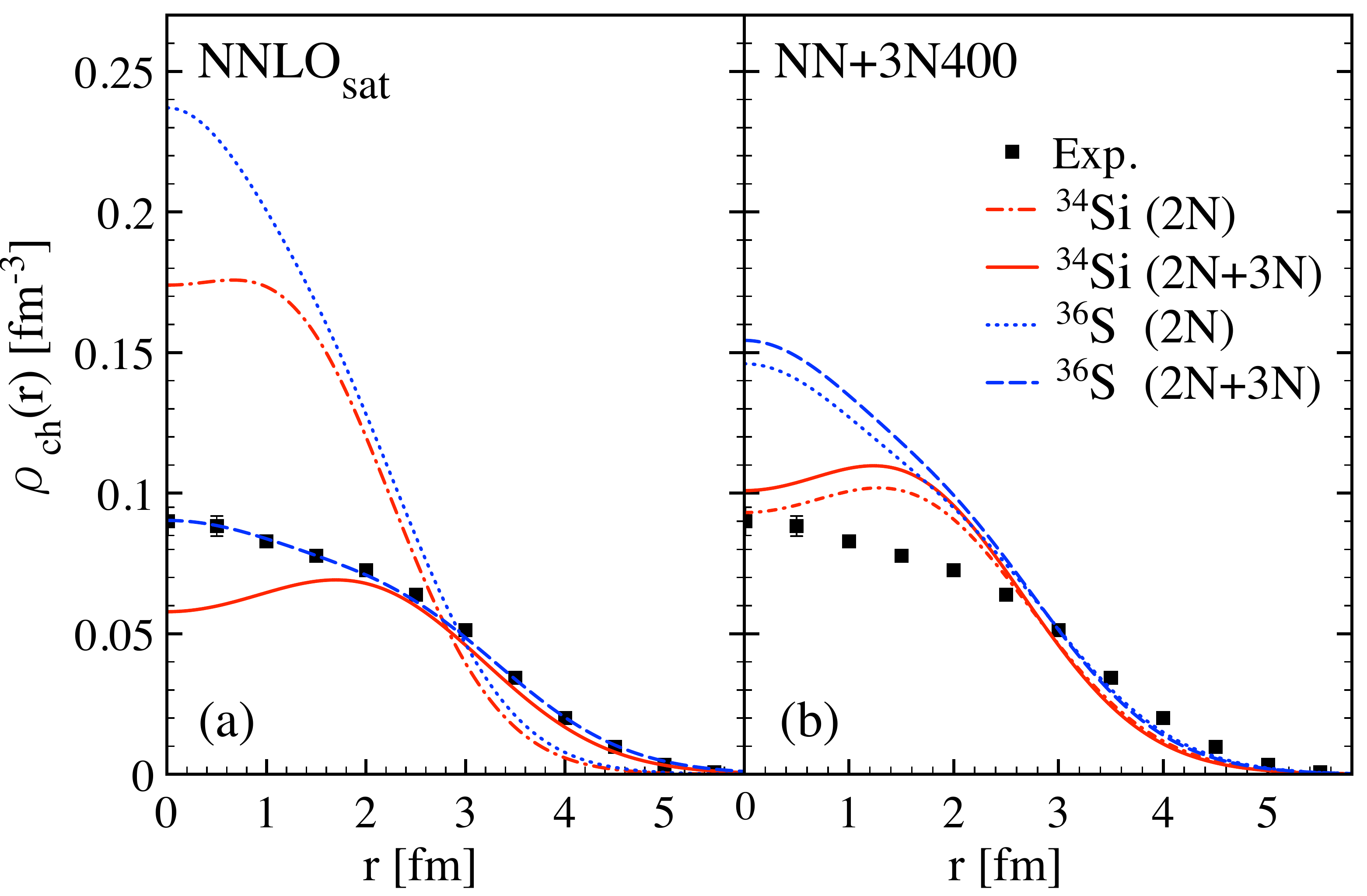}
\caption{(Color online) Charge density distributions of $^{34}$Si and $^{36}$S computed at the ADC(2) level with and without 3N interactions. (a) NNLO$_{\text{sat}}$ Hamiltonian. (b) NN+3N400(1.88) Hamiltonian. The experimental charge density of $^{36}$S is also shown~\cite{Rychel83}.}
\label{3NF}
\end{figure}

To single out the effect of the 3N interaction, charge density distributions of $^{34}$Si and $^{36}$S computed at the ADC(2) level with and without it are displayed in Fig.~\ref{3NF} for both NNLO$_{\text{sat}}$ and NN+3N400(1.88) Hamiltonians.

For NNLO$_{\text{sat}}$, the inclusion of the 3N interaction has a severe impact on the charge density distribution and on the rms charge radius of $^{36}$S, which is eventually in good agreement with data as already discussed above. Generically, the charge density is strongly reduced in the bulk and pushed towards the outside. The maximum of the density in $^{34}$Si moves down and away from the center, i.e. from about $1$\,fm to about $2$\,fm, enlarging radially the bubble structure. At the same time, the 3N force increases significantly the depletion in the center. Given that the charge density overall shrinks in the bulk, this leads to a pronounced enlargement of the F factor from $0.01$ with 2N interaction only to $0.15$ when the 3N interaction is incorporated. 

At first sight, the effect of the 3N interaction, which first enters at NNLO in Weinberg's power counting~\cite{Bedaque:2002mn,Nogga:2005vn,Epelbaum:2008ga}, seems anomalously large compared to the 2N only result. To investigate this point thoroughly, one would need to perform the calculation at various orders in the $\chi$-EFT power counting, not just remove the NNLO 3N contribution from the NNLO$_{\text{sat}}$ Hamiltonian as done here. If a strong impact of the NNLO contributions to the Hamiltonian were to be confirmed, it would additionally question the present use of a leading-order approximation to the charge-density operator\footnote{Independently of that, it is anyway necessary in principle to use consistent Hamiltonian and charge operators, e.g. see Ref.~\cite{Krebs:2016rqz}.}. While being beyond the scope of the present paper, these issues need to be investigated in the future.

\begin{table}
\centering
\begin{tabular}{|c|c||c|}
\hline
$^{34}$Si & & $F_{{\rm ch}}$  \\
\hline
\hline
NNLO$_{\text{sat}}$ & \,\,2N  & \,\,\,\,0.01\,\,\,\, \\
  			NNLO$_{\text{sat}}$		& \,\,2N+3N\,\, & 0.15 \\
\hline
\,\,NN+3N400(1.88)\,\, & \,\,2N & 0.08 \\
\,\,NN+3N400(1.88)\,\,		 & \,\,2N+3N\,\, & 0.08 \\
\hline
\end{tabular}
\caption{Charge depletion factors in $^{34}$Si computed at the ADC(2) level with and without 3N forces for NNLO$_{\text{sat}}$ and NN+3N400(1.88).}
\label{2N3NF}
\end{table}

For NN+3N400(1.88), the impact of the 3N interaction is much weaker. In $^{36}$S, the 3N interaction only slightly pushes the charge density away from the center, which is not sufficient to bring the charge radius in agreement with data. The charge density of $^{34}$Si is essentially untouched and so is the bubble structure. Independently of the inclusion of the 3N force, the F factor remains moderate and equal to $0.08$. 

The present analysis demonstrates that, as expected on general grounds, the effect of the 3N interaction depends of the Hamiltonian, i.e. it is a function of the partner 2N interaction and of the overall quality of the Hamiltonian it is part of. When considering the overall (2N+3N) Hamiltonian, only the latter remains and the variability of the results is more transparent and systematic, as discussed in Sec.~\ref{choiceH}.

\subsection{Comparison to SM and MR-EDF calculations}

We wish to compare results of ab initio SCGF calculations based on the NNLO$_{\text{sat}}$ Hamiltonian to those obtained from alternative theoretical methods. As EDF calculations of Refs.~\cite{Yao12,Yao13} demonstrated, correlations beyond any Hartree-Fock or static SR-EDF method significantly affect potential bubble structures. Consequently, we focus on MR-EDF calculations that do incorporate long-range correlations explicitly. Table.~\ref{F34SiEDF} compares proton and charge depletion factors obtained in $^{34}$Si from present ab initio SCGF calculations to those obtained from state-of-the-art MR-EDF~\cite{Yao12,Yao13} and SM calculations. It is to be noticed that the experimental charge density of $^{36}$S is equally well reproduced by ab initio SCGF and MR-EDF calculations. The reproduction is fair in the SM calculation, knowing that the charge density is computed by weighting one-body wave-functions derived from a well tailored Woods-Saxon potential with the single-particle occupations  obtained from the SM calculation.

Multi-reference EDF calculations of Refs.~\cite{Yao12} and~\cite{Yao13} are based on non-relativistic and relativistic energy functionals, respectively. Both sets of calculations include long-range correlations associated with the restoration of $U(1)$ and $SO(3)$ symmetries, i.e. with the restoration of good nucleon numbers and angular momentum, as well as with the (large amplitude) fluctuation of the intrinsic axial quadrupole deformation. The calculations are benchmarked against the known spectroscopy of $^{34}$Si (i.e. $E_{2^+_1}$, $E_{0_2^+}$, $B(E2; 0^+_1\rightarrow 2^+_1)$ and $\rho(E0; 0^+_1\rightarrow 0^+_2)$). While the agreement is satisfactory in both cases, the calculations based on a the relativistic point-coupling PC-PK1 parameterization~\cite{zhao10a}, augmented with a contact interaction to treat pairing correlations within the BCS approximation, are particularly performing and display an appropriate amount of intrinsic shape mixing in the low-lying states of interest. Shell-model calculations of Ref.~\cite{Grasso09} employ the USD interaction~\cite{brown88a} in the sd valence space. As such, long-range correlations are included via the full diagonalization of the effective interaction in the valence space. A good reproduction of $2^+_1$ and $0_2^+$ states in $^{34}$Si is however likely to necessitate 2p-2h intruders outside the sd shell such that the benchmarking against the known spectroscopy of $^{34}$Si is out of the reach of sd SM calculations~\cite{brown16a}. Similarly, the excitation energy of $2^+_1$ and $0_2^+$ states in $^{34}$Si is not yet available in ab initio SCGF calculations for benchmarking. However, SCGF calculations can be tested against spectra of neighboring nuclei accessed via, e.g., one-nucleon addition and removal experiments. This is postponed to the next section.

\begin{table}
\centering
\begin{tabular}{|c||c|c|c|c|c|}
\hline
$^{34}$Si & SCGF & SCGF* & MREDF~\cite{Yao12}  & MREDF~\cite{Yao13} & SM~\cite{Grasso09} \\
\hline
\hline
$F_p$ 			& 0.34 & 0.34 & 0.21 & 0.22  & 0.41 \\
$F_{{\rm ch}}$ 	& 0.15 & 0.19* & 0.09 & 0.11  & 0.28 \\
\hline
\end{tabular}
\caption{Point-proton and charge depletion factors in $^{34}$Si from ab initio ADC(3) SCGF calculations based on the NNLO$_{\text{sat}}$ Hamiltonian as well as from MR-EDF~\cite{Yao12,Yao13} and SM~\cite{Grasso09} calculations. In Refs.~\cite{Yao12,Yao13}, the charge density is obtained by folding the point proton density with a single gaussian characterized by an effective proton rms radius $R^{\text{eff}}_p = 0.8$ fm. For the sake of proceeding to a meaningful comparison we also report SCGF results obtained following this procedure and denote them as SCGF*.}
\label{F34SiEDF}
\end{table}

As seen in Tab.~\ref{F34SiEDF}, charge depletion factors of $^{34}$Si obtained from both MR-EDF calculations are essentially identical and predicted to be around $0.10$. The depletion factor is predicted to be  larger in ab initio calculations ($0.19$*) and significantly larger in SM calculations ($0.28$), with the caveat that the charge density is obtained through a rather had hoc procedure in the latter case. Interestingly, the difference between SCGF and MR-EDF calculations essentially originates from the underlying spherical mean-fields. While present ADC(1) calculation predicts $F_{{\rm ch}}=0.31$*, the charge depletion factor is $0.21$ in the SR-EDF calculation based on a spherical Hartree-Fock reference state~\cite{Yao12}. Adding long-range correlations, the suppression of the bubble structure is identical in both sets of calculations in spite of the fact that the many-body schemes employed to do so are very different. This may reflect the need for EDF parameterizations to be fitted at the MR level, i.e. once long-range correlations are included. All in all, predictions from present ab initio calculations and from SM calculations leave more hope to observe a bubble structure in the charge density of $^{34}$Si than present-day MR-EDF calculations.

\section{Spectroscopy}
\label{spectro}

\subsection{One-nucleon addition and removal spectra}
\label{subsec_sep_E}

The spectral strength distribution displays one-nucleon separation energies 
\begin{equation}
E^{\pm}_k \equiv \pm \big( E^{\text{A} \pm 1}_k - E^{\text{A}}_0 \big) \label{sepEnergies}
\end{equation}
against spectroscopic factors
\begin{equation}
SF_{k}^{\pm} \equiv \sum_{p} S_{k}^{\pm pp} \, , \label{spectrofactor}
\end{equation}
for all final states of the $A\pm1$ systems reached by adding/removing one nucleon to/from the A-body ground-state of interest. Spectroscopic factors are computed from one-nucleon addition and removal spectroscopic probability matrices defined through
\begin{subequations}
\label{spectroproba}
\begin{eqnarray}
S_{k}^{+pq} &\equiv&  \langle \Psi^{\text{A}}_{0} |  a_p | \Psi^{\text{A+1}}_{k} \rangle \langle \Psi^{\text{A+1}}_{k} | a^\dagger_q | \Psi^{\text{A}}_{0} \rangle \, \, \, , \label{spectroprobaplus} \\
S_{k}^{-pq} &\equiv& \langle \Psi^{\text{A}}_{0} | a^\dagger_q | \Psi^{\text{A-1}}_{k} \rangle \langle \Psi^{\text{A-1}}_{k} | a_p | \Psi^{\text{A}}_{0} \rangle    \, \, \, . \label{spectroprobamoins}
\end{eqnarray}
\end{subequations}

\begin{figure}
\centering
\includegraphics[width=9.0cm]{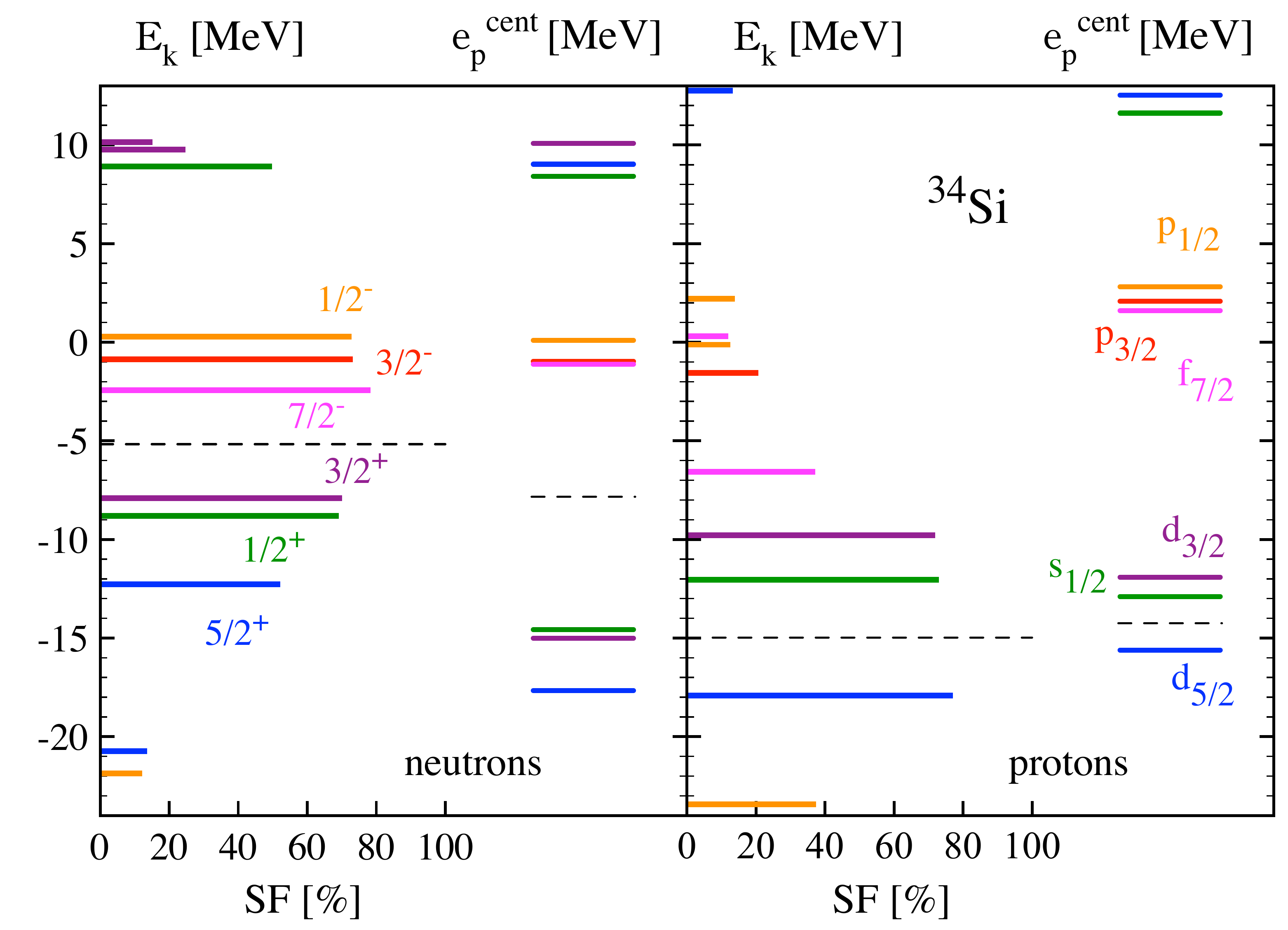}
\caption{(Color online) One-nucleon addition and removal spectral strength distribution along with associated effective single-particle energies in $^{34}$Si. Left panel: neutrons. Right panel: protons. Dashed lines denote the Fermi energies and separate the addition and removal parts of the spectra.}
\label{spectral_34Si}
\end{figure}

\begin{figure}
\centering
\includegraphics[width=9.0cm]{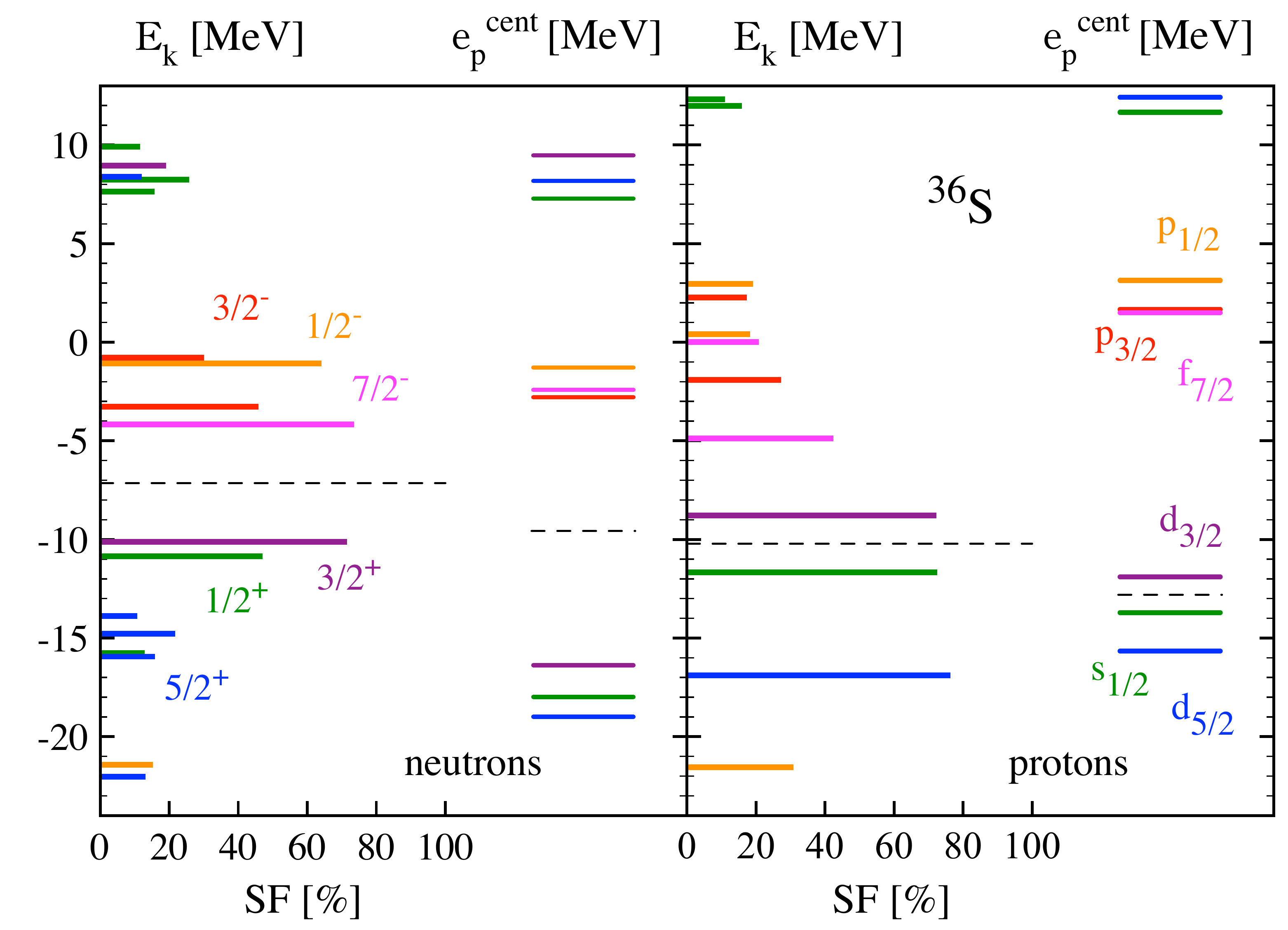}
\caption{Same as Fig.~\ref{spectral_34Si} for $^{36}$S.}\label{spectral_36S}
\end{figure}

Self-consistent Green's function calculations of $^{34}$Si and $^{36}$S ground states automatically access the information on neighboring $A\pm1$ systems associated with Eqs.~\ref{sepEnergies}-\ref{spectroproba}. For reference, one-neutron and one-proton addition and removal spectral strength distributions associated with the ground state of $^{34}$Si ($^{36}$S) and calculated at the ADC(3) level on the basis of the  NNLO$_{\text{sat}}$ Hamiltonian are displayed over a wide energy range in Fig.~\ref{spectral_34Si} (Fig.~\ref{spectral_36S}). Bars above (below) the dashed line denote states in the nucleus with one nucleon more (less).
The length of the bars characterizes the fragmentation of the strength in the present many-body calculation.

\subsection{Comparison to experimental data}
\label{comp_data}

\begin{figure}
\centering
\includegraphics[width=9.0cm]{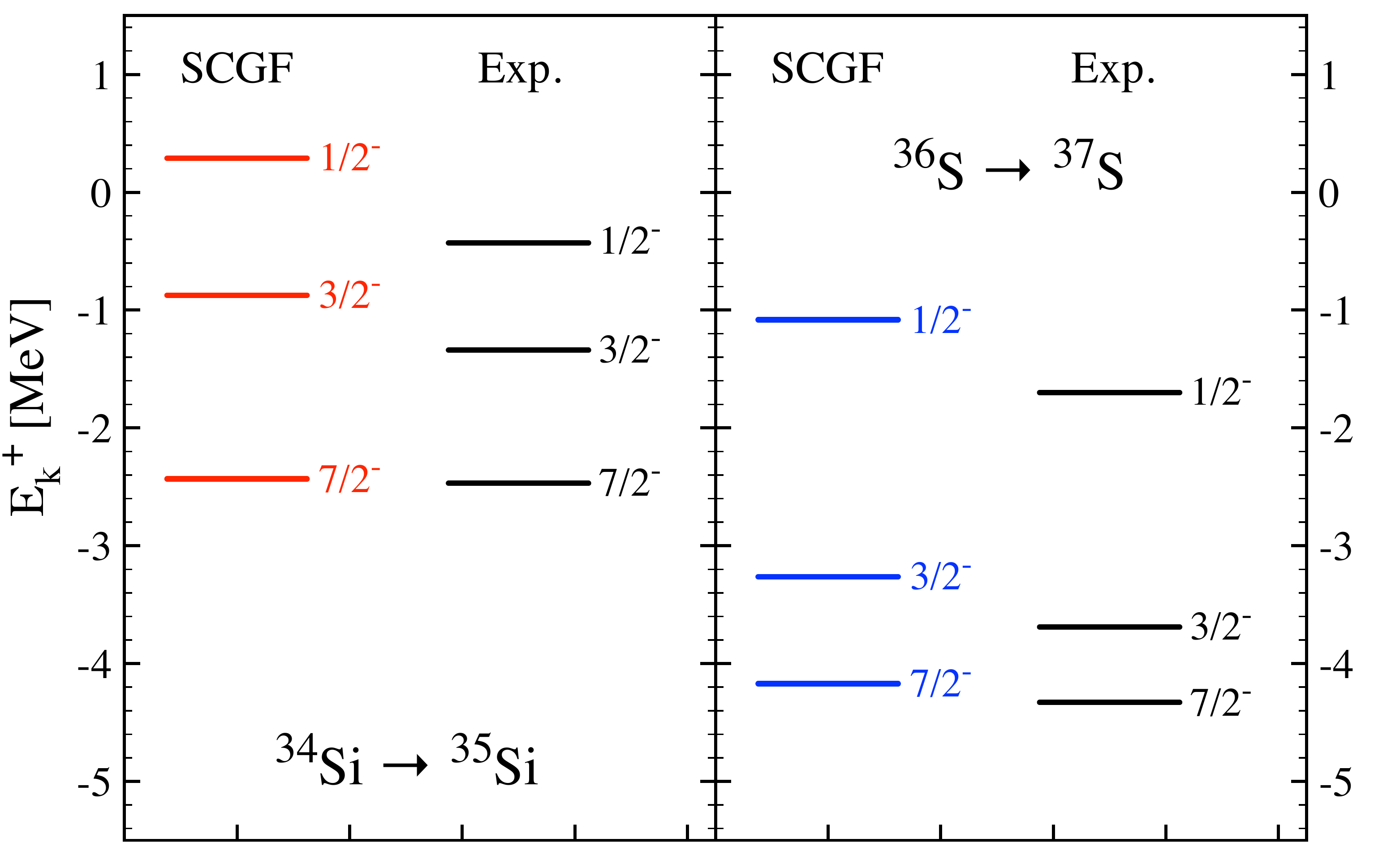}
\caption{(Color online) One-neutron addition energies to the lowest-lying states above the Fermi energy from ADC(3) SCGF calculations based on the  NNLO$_{\text{sat}}$ Hamiltonian. Left panel: $^{34}$Si (final states in $^{35}$Si). Right panel: $^{36}$S (final states in $^{37}$S). Experimental energies obtained via (d,p) reactions are taken from Refs.~\cite{Thorn84, Eckle89, Burgunder14}.}
\label{spectral_exp}
\end{figure}

To better typify present theoretical predictions and compare them to available experimental data, one-neutron additional energies to the lowest-lying states of $^{35}$Si and $^{37}$S ($N=21$) with dominant strength are shown in Fig.~\ref{spectral_exp} against experimental data obtained from (d,p) reactions on $^{34}$Si~\cite{Burgunder14} and $^{36}$S~\cite{Thorn84, Eckle89}.
One notices that the theoretical ordering of the $7/2^-$, $3/2^-$ and $1/2^-$ states is correct in $^{35}$Si and $^{37}$S and the distance between the states is in fair agreement with data. Even though the theoretical spectra are slightly too diluted, they are incomparably better than with the NN+3N400 Hamiltonian for which they come out much more spread out  (see Ref.~\cite{Papuga:2014hsa} for a typical example in K isotopes). Finally, the three lowest-lying separation energies are also well reproduced on an absolute scale, which relates to the fact that the error on the separation energy between the ground states of $^{34}$Si ($^{36}$S) and $^{35}$Si ($^{37}$S) is only $40$\,keV ($160$\,keV).  While the 3N interaction is key to obtain the correct density of states, many-body correlations are essential to position the spectrum on an absolute scale, e.g. going from ADC(2) to ADC(3) lowers $E^{+}_{7/2^-}$ and $E^{+}_{3/2^-}$ ($E^{+}_{1/2^-}$) by slightly more (less) than $1$\,MeV in the direction of experimental data in $^{35}$Si. In view of this, we can speculate that missing ADC(4) correlations might shift one-nucleon separation energies of dominant peaks on a level of $100$\,keV. Overall, the largest disagreement with data relates to the $1/2^-$ state in $^{35}$Si that is wrongly predicted unbound by $300$\,keV, i.e. $700$\,keV above the experimental state. In addition to the possible improvement brought about by a better $\chi$EFT Hamiltonian and by the inclusion of ADC(4) correlations, this low angular-momentum state near the continuum threshold is probably significantly impacted by the use of a truncated HO basis to expand the many-body Schr\"odinger equation. The residual sensitivity of this state to the value of ($N_{\text{{\rm max}}},\hbar\omega$) testifies that it is less well converged than the other low-lying states and should probably lie few hundred keV lower.

\begin{figure}
\centering
\includegraphics[width=9.0cm]{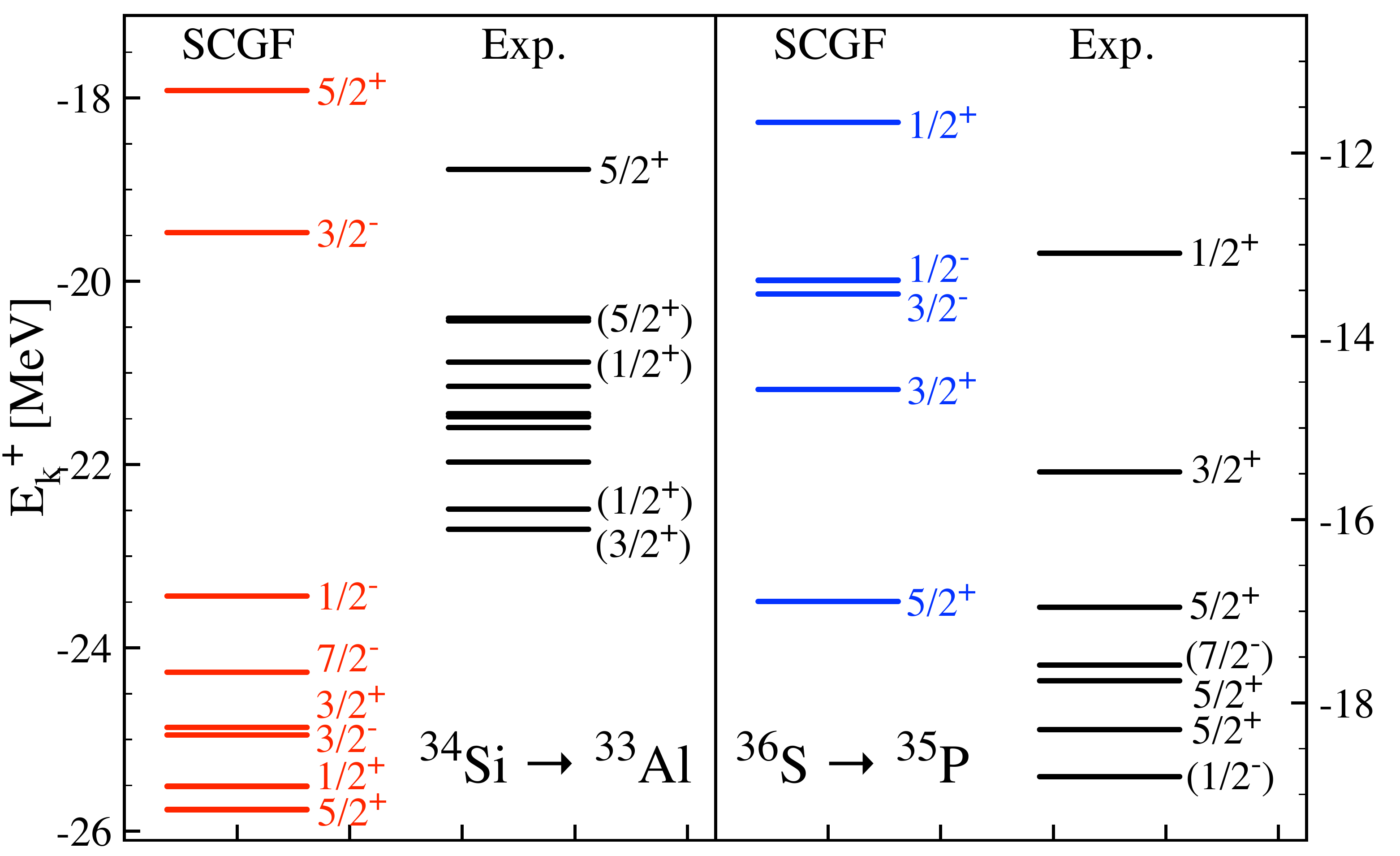}
\caption{(Color online) One-proton removal energies to the lowest-lying states below the Fermi energy from ADC(3) SCGF calculations based on the  NNLO$_{\text{sat}}$ Hamiltonian. Left panel: $^{34}$Si (final states in $^{33}$Al). Right panel: $^{36}$S (final states in $^{35}$P). Experimental energies obtained via knock-out reactions on $^{34}$Si ($^{36}$S) are taken from Ref.~\cite{mutschler16b} (Refs.~\cite{khan85a,mutschler16a}). In both cases, the larger the excitation energy of a given state, the more negative the associated one-proton removal energy.}
\label{spectral_exp_-1}
\end{figure}

Similarly, one can analyse one-proton removal energies to the lowest-lying states of $^{33}$Al ($Z=13$) and $^{35}$P ($Z=15$) as obtained from knock-out experiments on $^{34}$Si~\cite{mutschler16b} and $^{36}$S~\cite{khan85a,mutschler16a}. The comparison is shown in Fig.~\ref{spectral_exp_-1}.
In $^{35}$P, the sequence of $1/2^{+}$, $3/2^{+}$ and $5/2^{+}$ is qualitatively consistent with data, which is a key feature with regards to the occurrence of the proton bubble. Quantitatively though, the separation energy to the $1/2^{+}$ ground state is $1.5$\,MeV too small whereas excitation energies of the $3/2^{+}$ and $5/2^{+}$ states are $0.5$\,MeV and $1.5$\,MeV too large, respectively. The theoretical spectrum displays additional low-lying $1/2^{-}$ and $3/2^{-}$ states with small spectroscopic strength. However, these small fragments are unlikely to be fully converged with respect to the many-body truncations even at the ADC(3) level.

In $^{33}$Al, no state except for the $5/2^{+}$ ground-state has firm spin-parity assignment. The ground-state spin-parity is reproduced in the calculation although the corresponding separation energy is $1.5$\,MeV too small. The experimental spectrum appears much denser, i.e. more fragmented, between $0$ and $4$\,MeV excitation energy than in $^{35}$P. It is not what is obtained theoretically where only two states arise in this energy window. While at least two tentative $1/2^{+}$ states are seen experimentally below $5$\,MeV with small cross sections, the first calculated $1/2^{+}$ state is located at about $7.5$\,MeV. This situation most probably reflects that $^{33}$Al is on the edge of the so-called island of inversion. Capturing the strong associated quadrupole-quadrupole correlations is probably beyond the reach of our many-body truncation scheme. As a matter of fact, improving our description of the low-lying spectroscopy of $^{33}$Al and/or calculating its ground-state quadrupole moment~\cite{Heylen:2016nbh} constitutes a challenging task for theory. As for ab initio SCGF method, one possible way out consists in allowing for spherical symmetry to be spontaneously broken in the calculation. This would however require for good angular momentum to be eventually restored, which is yet to be formulated within the frame of the SCGF method.

All in all, the spectroscopy of nuclei obtained from $^{34}$Si and $^{36}$S by the addition (removal) of a neutron (proton) is in fair agreement with data at the present stage of development of ab initio calculations in mid-mass nuclei. Still, the significant disagreement with data in $^{33}$Al is at variance with the other cases and deserves more attention in the future. At this point in time, the comparison to such refined spectroscopic observables give further confidence into the predictions made about the occurrence of a bubble structure in $^{34}$Si. 

\subsection{Spin-orbit splitting and bubble structure}
\label{bubble_SO}

\begin{table}
\centering
\begin{tabular}{|c||c|c||c|}
\hline
$E_{1/2^-}\!-\!E_{3/2^-}$ & $^{37}$S &  $^{35}$Si & $^{37}$S$\rightarrow ^{35}$Si \\
\hline
\hline
SCGF & \,\, 2.18 \,\, & \,\,1.16 \,\, & \,\, -1.02 (-47$\%$) \,\, \\
(d,p) & 1.99 & 0.91 & -1.08 (-54$\%$) \\
\hline
\end{tabular}
\caption{Splitting $E_{1/2^-}\!-\!E_{3/2^-}$ in $^{37}$S and $^{35}$Si (in MeV) as obtained from ADC(3) SCGF calculations  based on the NNLO$_{\text{sat}}$ Hamiltonian and from (d,p) experiments~\cite{Thorn84, Eckle89, Burgunder14}.}
\label{splitting_E}
\end{table}

As alluded to in the introduction, it has been speculated that the presence of a semi-bubble in the center of a light nucleus could feedback on the splitting between low angular-momentum spin-orbit partners in the $A\pm1$ systems. Consequently, the question presently arises to which extent the appearance of a significant bubble ($F_{{\rm ch}}=0.15$) when removing two protons from $^{36}$S to $^{34}$Si impacts the size of the splitting between the low-lying $1/2^-$ and $3/2^-$ states when going from $^{37}$S to $^{35}$Si. As reported in Tab.~\ref{splitting_E}, this splitting is predicted to decrease from $2.18$\,MeV to $1.16$\,MeV, i.e. to be reduced by $1.02$\,MeV (47$\%$). This sudden reduction by about 50$\%$ is unique over the nuclear chart and is in good agreement with data both on an absolute and a relative scale.

\begin{figure}
\centering
\includegraphics[width=8.5cm]{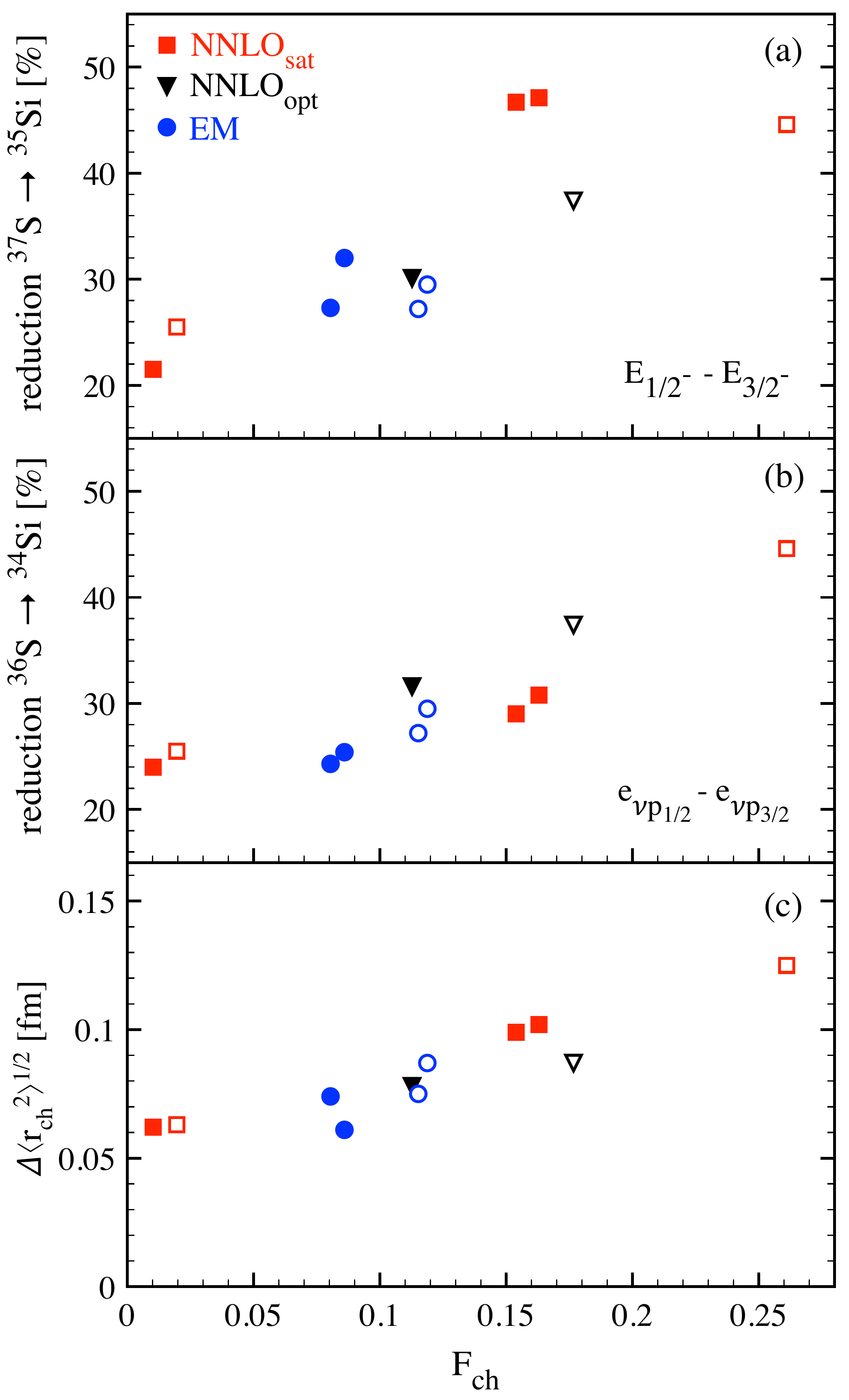}
\caption{(Color online) (a) Reduction of the $E_{1/2^-}\!-\!E_{3/2^-}$ spin-orbit splitting going from $^{37}$S to $^{35}$Si against the charge depletion factor $F_{{\rm ch}}$ in $^{34}$Si for all presently available SCGF calculations. (b) Same as the upper panel but for the neutron ESPE spin-orbit splitting $e_{1p_{1/2}}\!-\!e_{1p_{3/2}}$ going from $^{36}$S to $^{34}$Si. (c) Change of rms charge radius between  $^{36}$S and $^{34}$Si against the charge depletion factor $F_{{\rm ch}}$ of the latter for all presently available SCGF calculations. Open symbols correspond to ADC(1), i.e. HF, calculations.}
\label{correlationplot}
\end{figure}

To better establish whether there is an actual correlation between the appearance of the bubble in $^{34}$Si and the reduction of the $E_{1/2^-}\!-\!E_{3/2^-}$ splitting when going from $^{37}$S to $^{35}$Si, the analysis is first repeated with NN+3N400(1.88) for which the bubble structure is predicted to be less pronounced, i.e. $F_{{\rm ch}}=0.08$ instead of $F_{{\rm ch}}=0.15$ for NNLO$_{\text{sat}}$ at the ADC(2) level. The splitting between the $1/2^-$ and $3/2^-$ states only reduces by 27$\%$\footnote{The spectra being significantly more spread out with NN+3N400(1.88), it is not meaningful to compare the reduction of the splitting in absolute values. Still, let us mention that the splitting decreases by about $800$\,keV out of the $3$\,MeV value it takes in $^{37}$S, i.e. it actually decreases less in absolute value than with NNLO$_{\text{sat}}$ is spite of being $1$\,MeV larger (too large) in the first place.} when going from $^{37}$S to $^{35}$Si, which is indeed in proportion of $0.08/0.15$ of the reduction seen with  NNLO$_{\text{sat}}$. To establish the correlation on firmer a ground, the reduction of the $E_{1/2^-}\!-\!E_{3/2^-}$ splitting is plotted in the upper panel of Fig.~\ref{correlationplot} against the charge depletion factor $F_{{\rm ch}}$ in $^{34}$Si for all presently available SCGF calculations\footnote{The theoretical data set contains calculations performed with and without 3N interactions at the ADC(1), ADC(2) and ADC(3) levels for the four Hamiltonians under present consideration.}. This systematic gives quantitative credit to the existence of a correlation between the appearance of a bubble structure and the reduction of spin-orbit splittings of low-lying/low angular-momentum states. 

Last but not least, the lower panel of Fig.~\ref{correlationplot} highlights the fact that the depletion factor  of $^{34}$Si is significantly correlated with the difference of charge rms radius between $^{36}$S and $^{34}$Si. Furthermore, the sensitivity to the bubble leads to a variability of the order of $0.05$ fm between the case where there would be no bubble ($F_{{\rm ch}}=0$) in $^{34}$Si and our current prediction ($F_{{\rm ch}}=0.15$), which is several times larger than the typical 1$\sigma$ uncertainty for the measurement of absolute charge radii whenever doable~\cite{fricke04a}. The latter acts as a strong motivation to measure the absolute charge radius of $^{34}$Si or its isotopic shift with respect to $^{28,29,30}$Si whose charge radii are known~\cite{fricke04a}.

\subsection{Effective single-particle energies}
\label{subsec_espe}

The original speculation that the presence of a bubble could lead to a reduction of the splitting between spin-orbit partners characterized by moderate angular momenta~\cite{ToddRutel04} was based on a mean-field picture in which the effective one-body spin-orbit potential is essentially proportional to the derivative of the point-nucleon density distribution. We have seen above that this correlation does indeed manifest when looking at many-body energies in spite of the fact that no one-body spin-orbit potential proportional to the derivative of the density is explicitly at play in the ab initio resolution of the many-body Schr\"odinger equation. It is thus of interest to reverse engineer and study to which extent a consistent picture does indeed emerge in the one-body shell structure that underlines the correlated many-body system. It is indeed useful to {\it interpret} the behavior of observable one-nucleon separation energies in simple terms. One must however be clear that effective single-particle energies are not observable, i.e. the picture provided by this interpretation is only valid {\it within} the theoretical scheme used and must not be extrapolated a priori to other theoretical schemes\footnote{A theoretical scheme is fully defined by the given of interacting degrees of freedom and of the Hamiltonian that drives their dynamics, without any further freedom to perform unitary transformations of that Hamiltonian. Indeed, the value of effective single-particle energies can be changed within the theory through a unitary transformation without changing observable one-nucleon separation energies~\cite{Duguet15}. This demonstrates that both quantities possess different ontological status within the frame of many-body quantum mechanics and that a quantitative link between them can only be made {\it within} the theory by fixing the freedom allowed by unitary transformations.}~\cite{Duguet15}.

Having the complete set of one-nucleon addition/removal energies and of spectroscopic probability matrices at hand, the centroid Hamiltonian can be constructed to give access, via its diagonalization, to the one-nucleon shell structure associated with effective single particle energies (ESPEs)~\cite{baranger70a}. In the corresponding eigenbasis, ESPEs write as
\begin{equation}
e^{\text{cent}}_{p} = \sum\limits_{k \in {\cal H}_{A-1}} E^{-}_{k} \, S^{-pp}_{k} + \sum\limits_{k \in {\cal H}_{A+1}} E^{+}_{k} \, S^{+pp}_{k} \, . \label{ESPEs}
\end{equation}
The set of ESPEs computed for $^{34}$Si and $^{36}$S appear in Figs.~\ref{spectral_34Si} and~\ref{spectral_36S}. By comparing ESPEs to the associated spectral strength distribution, one appreciates the fragmentation of the latter. Correspondingly, any given ESPE is not in one-to-one correspondence with the dominant fragment of appropriate spin and parity but is a centroid of all the fragments with the same spin and parity. This is particularly striking in $^{36}$S where the ordering of neutron ESPEs above the Fermi level differs from the ordering of the main peaks 
in the spectral strength distribution of $^{37}$S. Indeed, the  $1f_{7/2}$ is above the $1p_{3/2}$ whereas the main $7/2^-$ fragment is below the main $3/2^-$ fragment. The orderings of both spectra become consistent when removing two protons given that the strength is less fragmented in $^{35}$Si than in $^{37}$S.

\begin{table}
\centering
\begin{tabular}{|c||c|c||c|}
\hline
$e^{\text{cent}}_{1p_{1/2}}\!-\!e^{\text{cent}}_{1p_{3/2}}$ & $^{36}$S &  $^{34}$Si & $^{36}$S$\rightarrow ^{34}$Si\\
\hline
\hline
SCGF & \,\, 1.50 \,\, & \,\, 1.07 \,\, & \,\, -0.43 (-29$\%$) \,\, \\
SM & - & - & -0.38 (-25$\%$) \\
\hline
\end{tabular}
\caption{ESPE spin-orbit splitting $e^{\text{cent}}_{1p_{1/2}}\!-\!e^{\text{cent}}_{1p_{3/2}}$ in $^{36}$S and $^{34}$Si (in MeV) as obtained from ADC(3) SCGF calculations based on the NNLO$_{\text{sat}}$ Hamiltonian and from sd-pf SM calculations~\cite{Burgunder14}.}
\label{splitting_e}
\end{table}

In connection with the study performed in Sec.~\ref{subsec_sep_E} on the splitting between low-lying $1/2^-$ and $3/2^-$ states reached by one-nucleon addition processes, we now focus on the evolution of the neutron 1p$_{3/2}$-1p$_{1/2}$ ESPE spin-orbit splitting when going from $^{36}$S to $^{34}$Si. As qualitatively visible in Figs.~\ref{spectral_34Si} and~\ref{spectral_36S}, and as quantitatively reported in Tab.~\ref{splitting_e}, the ESPE spin-orbit splitting reduces from $1.50$\,MeV to $1.07$\,MeV, i.e. it is lowered by $29\%$.  This feature, as well as the correlation this reduction of the ESPE splitting entertains with the charge depletion factor, is also illustrated in Fig.~\ref{correlationplot}.  

The reduction of the 1p$_{3/2}$-1p$_{1/2}$ ESPE spin-orbit splitting obtained from of a full sd-pf SM calculation~\cite{Burgunder14} is also reported in Tab.~\ref{splitting_e}. Within that theoretical scheme, which reproduces by construction the experimental energies of the main fragments in $^{37}$S to $^{35}$Si once full correlations are included, the ESPE spin-orbit splitting reduces by $25\%$, which is close to the reduction obtained within the SCGF calculation based on the NNLO$_{\text{sat}}$ Hamiltonian. While ESPEs do not have to agree, even if both theoretical calculations equally reproduce observable many-body energies, the reduction of the ESPE spin-orbit splitting happen to be similar in both schemes.

The analysis of the reduction of the ESPE splitting as a function of the depletion factor allows to better understand how the reduction of the observable many-body splitting $E_{1/2^-}\!-\!E_{3/2^-}$ operates in the present ab initio calculation. The $E_{1/2^-}\!-\!E_{3/2^-}$ splitting being a coherent combination of the ESPE contribution and of many-body correlations~\cite{Duguet15}, one sees that the former contribution is responsible for about $50\%$ of the total reduction while the latter generates the other $50\%$ in the full calculation with NNLO$_{\text{sat}}$. The contribution from many-body correlations relates to the fact that the $3/2^-$ strength is fragmented in $^{36}$S but not in $^{34}$Si. This is testified by the presence of the second $3/2^-$ state just above the $1/2^-$ state in the left panel of Fig.~\ref{spectral_36S}. As a result, the lowest-lying $3/2^-$ state is mechanically pushed away from the $1/2^-$ state, thus adding to the $E_{1/2^-}\!-\!E_{3/2^-}$ splitting beyond the sole migration of the ESPE centroids. As the bubble disappears and the charge radius decreases in the lowest two panels of Fig.~\ref{correlationplot}, both the relative migration of 1p$_{3/2}$ and 1p$_{1/2}$ ESPEs and the fragmentation of the $3/2^-$ strength (not shown here) diminish, thus leading to a less pronounced reduction of the $E_{1/2^-}\!-\!E_{3/2^-}$ splitting when going from $^{37}$S to $^{35}$Si.

\section{Conclusions}
\label{conclusions}

Semi-bubble or bubble structures have been invoked in connection with hypothetical super-heavy or hyper-heavy nuclei as a result of a collective quantum mechanical effect, further sustained by the compromise between the large repulsive Coulomb interaction and the strong force that binds them. In lighter systems, semi-bubble structures have been conjectured on the basis of the sole quantum mechanical effect, which finds its source in the sequence of occupied and unoccupied single-particle states near the Fermi energy in an independent-particle or a mean-field picture.

In connection with the latter category, we have studied the potential bubble $^{34}$Si nucleus on the basis of state-of-the-art ab initio quantum many-body methods. The occurrence of a depletion in the center of the charge density distribution and its correlation with an anomalously weak splitting between low angular-momentum spin-orbit partners was originally postulated on the basis of a mean-field rationale. It was thus of interest to investigate this speculation on the basis of fully correlated solution of the many-body Schr\"odinger equation.

We have performed ab initio self-consistent Green's function many-body calculations of $^{34}$Si and $^{36}$S. The focus was put on binding energies, charge density distributions and charge root mean square radii of $^{34}$Si and $^{36}$S as well as on the low-lying spectroscopy of nuclei obtained via the addition (removal) of a neutron (proton). Predictions regarding the (non)existence of the bubble structure in $^{34}$Si have been shown to vary significantly with the input chiral effective field theory Hamiltonian. However, demanding that the experimental charge density distribution and the root mean square radius of $^{36}$S, as well as $^{34}$Si and $^{36}$S  binding energies, are well reproduced has only left the NNLO$_{\text{sat}}$ Hamiltonian as a reliable candidate to perform these predictions. Additionally, the role of the three-nucleon interaction in the generation of the bubble structure has been scrutinized. 

Employing the NNLO$_{\text{sat}}$ Hamiltonian, a bubble structure has been convincingly predicted in $^{34}$Si. In particular, it has been demonstrated that the depletion in the center of the charge density distribution should be large enough to leave a visible signal in the form factor to be measured in a future electron scattering experiment. Furthermore, a clear correlation has been established between the occurrence of the bubble structure and the weakening of the $1/2^{-}\!-3/2^{-}$ splitting in the spectrum of $^{35}$Si. Consequently, the latter does offer an indirect hint for the existence of the bubble in $^{34}$Si~\cite{mutschler16b}. The interpretation in terms of the underlying one-nucleon shell structure has also been provided. 

Present results will have to be revisited with future generations of $\chi$EFT(-based) interactions. Indeed, the use of binding energies and radii of nuclei containing about $20$ nucleons to adjust the low-energy constants of two- and three-body interactions entering the NNLO$_{\text{sat}}$ Hamiltonian leaves many questions open regarding epistemology aspects of ab initio calculations and their associated predictive power.

Present results provide a strong motivation to measure the charge density distribution of $^{34}$Si in future electron scattering experiments on unstable nuclei. It was also shown that the measurement of its charge radius would already bring key insight. Furthermore, it is of interest to perform one-neutron removal experiments on $^{34}$Si and $^{36}$S in order to further test our theoretical spectral strength distributions over a wide energy range~\cite{mutschlerthesis}. Last but not least, other bubble nuclei candidates, possibly in excited states, should be  investigated theoretically and experimentally in the future.

\section*{Acknowledgements}

The authors warmly thank B. Bally, M. Bender, A. Brown, T. Cocolios, G. Hagen, V. Lapoux, M. Martini, W. Nazarewicz, G. Neyens, R. F. Garcia Ruiz, O. Sorlin and P. van Duppen for useful discussions and comments. Calculations were performed by using HPC resources from GENCI-TGCC (Contract No. 2016-057392) and at the DiRAC Complexity system at the University of Leicester (BIS National E-infrastructure capital grant No. ST/K000373/1 and STFC grant No. ST/K0003259/1). This work was supported by the United Kingdom Science and Technology Facilities Council (STFC) under Grant No. ST/L005816/1 and in part by the NSERC Grant No. SAPIN-2016-00033. TRIUMF receives federal funding via a contribution agreement with the National Research Council of Canada.

\bibliography{bubble}

\end{document}